\documentclass{aa} 
\usepackage{natbib,graphicx,amssymb}
\usepackage{dcolumn}

\begin{document}

\title{The circumstellar envelope of the C-rich post-AGB star
  HD~56126\thanks{based on observations taken at the European Southern
    Observatory, La Silla, Chile and observation obtained with ISO, an
    ESA project with instruments funded by ESA Member states
    (especially the PI countries: France, Germany, the Netherlands and
    the United Kingdom) with the participation of ISAS and NASA}}

\authorrunning{Hony et al.}

\titlerunning{The circumstellar envelope of HD~56126}

\author{
  S. Hony\inst{1,2}
  , A. G. G. M. Tielens\inst{3,4}
  , L. B. F. M. Waters\inst{1,5}
  , A. de Koter\inst{1}
}

\institute{
  Astronomical Institute ``Anton Pannekoek'', Kruislaan 403, 1098 SJ
  Amsterdam, The Netherlands
  \and
  RSSD-ESA/ESTEC, PO Box 299, 2200 AG Noordwijk, The Netherlands
  \and
  SRON Laboratory for Space Research Groningen, PO Box 800, 9700 AV
  Groningen, The Netherlands
  \and
  Kapteyn Astronomical Institute PO Box 800, 9700 AV
  Groningen, The Netherlands
  \and
  Instituut voor Sterrenkunde, K.U. Leuven, Celestijnenlaan 200B, 3001
  Heverlee, Belgium
}

\offprints{S.Hony, \email{shony@rssd.esa.int}}

\date{received \today; accepted date}

\abstract{ We present a detailed study of the circumstellar envelope
  of the post-asymptotic giant branch ``21~$\mu$m object'' HD~56126.
  We build a detailed dust radiative transfer model of the
  circumstellar envelope in order to derive the dust composition and
  mass, and the mass-loss history of the star. To model the emission
  of the dust we use amorphous carbon, hydrogenated amorphous carbon,
  magnesium sulfide and titanium carbide. We present a detailed
  parametrisation of the optical properties of hydrogenated amorphous
  carbon as a function of H/C content. The mid-infrared imaging and
  spectroscopy is best reproduced by a single dust shell from 1.2 to
  2.6$^{\prime\prime}$ radius around the central star.  This shell
  originates from a short period during which the mass-loss rate
  exceeded 10$^{-4}$~M$_{\odot}$/yr.  We find that the strength of the
  ``21''~$\mu$m feature poses a problem for the TiC identification.
  The low abundance of Ti requires very high absorption cross-sections
  in the ultraviolet and visible wavelength range to explain the
  strength of the feature. Other nano-crystalline metal carbides
  should be considered as well. We find that hydrogenated amorphous
  carbon in radiative equilibrium with the local radiation field does
  not reach a high enough temperature to explain the strength of the
  3.3$-$3.4 and 6$-$9~$\mu$m hydrocarbon features relative to the
  11$-$17~$\mu$m hydrocarbon features. We propose that the carriers of
  these hydrocarbon features are not in radiative equilibrium but are
  transiently heated to high temperature. We find that 2 per cent of
  the dust mass is required to explain the strength of the
  ``30''~$\mu$m feature, which fits well within the measured
  atmospheric abundance of Mg and S. This further strengthens the MgS
  identification of the ``30''~$\mu$m feature.  
  }

\maketitle

\section{Introduction}
\label{hd56:sec:intro}
Post-asymptotic giant branch stars (post-AGBs) are objects in which
the strong AGB mass-loss has ceased while the remaining star is not
(yet) hot enough to ionise the surrounding material. These objects are
characterised by a double-peaked spectral energy distribution (SED).
In the ultraviolet (UV) and visible range the central star is visible.
In the mid-infrared (IR) range the light is dominated by emission from
the dusty circumstellar envelope (CSE). Post-AGB stars are the prime
objects to study stellar evolution near the very end of the AGB
because they are the most recent evolutionary descendants of the AGB
stars. Mass loss near the tip of the AGB is not well understood. There
are indications that the mass-loss rates near the end of the AGB
exceed the values predicted by the existing dust driven wind models
(e.g. {CRL 2688}). There is also increasing evidence that
non-spherical mass loss is common near the end of the AGB phase. Since
AGB evolution timescales are determined by mass loss, these poorly
understood phenomena limit our knowledge of the evolution of these
objects and the subsequent stages.

Mass loss determinations are directly coupled to the composition of
the newly condensed dust.  The dust composition of the CSE reflects
the physical conditions in the outer atmospheres of the AGB star
during the phase when the dust was formed and the processing that took
place afterwards. We study the composition of the dust surrounding the
post-AGB star HD~56126.  This star is member of a class of rare
objects that exhibit an emission feature at 21~$\mu$m
\citep{1989ApJ...345L..51K}. The 12 \citep[see][ and references
therein]{1999IAUS..191..297K} post-AGB stars that show this feature
are commonly referred to as the ``21~$\mu$m objects''. The ``21~$\mu$m
objects'' form a group with rather homogeneous properties:
carbon-rich, low metallicity and s{\textendash}process enhanced
\citep{2000AnA...354..135V} with large IR excesses
\citep{1989ApJ...345L..51K}, spectral types G or F
\citep{1995ApJ...438..341H}. \citet{1999ApJ...516L..99V} have shown
that the shape of the ``21''~$\mu$m feature between different sources
shows hardly any variation. Recently we have reported the detection of
the same feature in two PNe \citep{2001A&A...378L..41H} and
\citet{Volk_Canberra} have reported on its detection in the PN IC~418.

Besides the ``21''~$\mu$m feature the ``21~$\mu$m objects'' exhibit
other infrared emission features that are unique to these
environments. The carriers of many of these features have not been
firmly identified and there is considerable debate in the literature
on the composition of the dust around the ``21~$\mu$m objects''. In
the next paragraphs we shortly review these features and the proposed
carriers.

The carrier of the ``21''~$\mu$m feature has remained unidentified for
more than a decade, although several candidates have been considered:
urea OC(NH$_2$)$_2$ \citep{1992AnA...254L...1S}, SiS$_2$
\citep{1993AnA...278..226G,1996ApJ...464L.195B}, fulleranes
\citep{1995MNRAS.277.1555W}, nano-diamond \citep{1998AnA...336L..41H},
hydrogenated amorphous carbon (HAC) \citep{2001ApJ...558L.129G} and
oxygen bearing side groups in coal \citep{2000AnA...362L...9P}.
However, none of the spectroscopic comparisons between the laboratory
spectra of the suggested materials and the emission from the post-AGB
stars is satisfactory. Recently we have suggested titanium carbide
nano-crystals (nano-TiC) as the carrier of the ``21''~$\mu$m feature
\citep{2000Sci...288..313V}. This identification is based on {\it i})
the excellent spectroscopic match of nano-TiC with the 21~$\mu$m
feature, {\it ii}) the fact that the nano-TiC resonance does not
depend on the crystal size for the size-range for which measurements
are available -- in good agreement with its observed constant profile,
{\it iii}) the fact that TiC is found in presolar grains
\citep{1996ApJ...472..760B} for which the isotopic ratios imply an
origin in the CSE of carbon-rich evolved stars
\citep{1993Metic..28..490A}.

We note that the nano-TiC identification is not commonly accepted
either.  The main counter argument is that the conditions (high
density and pressure during dust formation) required to form TiC in
the ejecta of carbon-rich evolved stars may not be met
\citep{2002ApJ...573..720K}.  Another argument against nano-TiC is the
strength of the ``21''~$\mu$m feature. Since Ti is a rare element, it
may be difficult for TiC to produce a feature that stands out as much
as it does.

As explained above the ``21~$\mu$m objects'' exhibit other emission
features for which the carriers are still under debate. Besides the
well-known unidentified infrared (UIR) bands at 3.3, 6.2, ``7.7'' and
11.2, generally attributed to polycyclic aromatic hydrocarbon (PAH)
molecules \citep{ATB89}, these sources exhibit a 3.4~$\mu$m feature
and broad emission features (so-called plateau features) between 6$-$9
and 11$-$17~$\mu$m \citep{1990ApJ...365L..23B,1996AnA...309..612J}. As
a class, the emission bands observed from the ``21~$\mu$m objects''
are distinctly different from the UIR features from most sources (see
Sect.~\ref{hd56:sec:iso-spectroscopy}).  \citet{2001ApJ...554L..87K}
have discussed these features in terms of alkane and alkene side
groups on very large aromatic molecules and small carbonaceous
particles.  \citet{1996ApJ...464..810G} have compared them to those
measured in coal (a natural terrestrial carbonaceous compound).  The
spectroscopic match with coal is very encouraging.  However, a
fundamental questions concerning the identification is whether
carbonaceous grains can reach the temperatures required to emit
strongly in the 6$-$9~$\mu$m range.

Lastly, these sources exhibit a strong, broad feature between
23$-$45~$\mu$m
\citep{1995ApJ...454..819O,1997AnA...317..859S,2000ApJ...535..275H},
called the ``30''~$\mu$m feature.  The carrier of the ``30''~$\mu$m
feature is identified with magnesium sulfide (MgS) \citep[See][ and
references therein]{2002A&A...390..533H}. The identification of the
``30''~$\mu$m feature with MgS seems solid but open questions remain
on the heating of the MgS grains and the amount of MgS required to
explain the prominence of the feature.

A meaningful approach towards the identification of the various dust
components in the CSE of these post-AGBs is that of radiative transfer
modelling of the SED of these objects. This may also help to improve
our understanding of the mass loss at the very end of the AGB phase.
Such modelling allows a comparison of the emission characteristics of
candidate materials and CSE emission and yields insight into the
relative contributions of the different materials present in the CSE.

We have elected to model {HD 56126} ({IRAS 07134+1005}, {SAO 96709},
{BD+10 1470}) in detail. HD~56126 has the strongest ``21''~$\mu$m
feature relative to the other dust features of all known ``21~$\mu$m
objects''. Therefore, HD~56126 provides the strongest constraint on
the abundances needed to explain the ``21''~$\mu$m feature.  This is
even more so because the photosphere of the star is metal poor
\citep{2000AnA...354..135V}. Both the central star and the dust shell
of this source have been well studied in the literature.  Especially
important in this respect are images at various wavelengths in the
mid-IR in which the source is well resolved \citep[][ and this
work]{1997ApJ...482..897M,1998ApJ...492..603D,2000ApJ...544L.141J,2002ApJ...573..720K}.
These images are essential to constrain the model parameters.

This paper is organised as follows: In
Sect.~\ref{hd56:sec:stellarparameters} we give the stellar parameters
of HD~56126. In Sect.~\ref{hd56:sec:observations} we present the
infrared measurements that we use. In Sect.~\ref{hd56:sec:basics} we
consider the energy balance in the nebula and constrain the stellar
effective temperature. In Sect.~\ref{hd56:sec:model} we list the
parameters that we use to model the absorption and emission in the CSE
of HD~56126. The optical properties of the dust constituents are
discussed in Sect.~\ref{hd56:sec:opticalprops}. In
Sect.~\ref{hd56:sec:model-results} we present the model results. In
Sect.~\ref{hd56:sec:discussion} we discuss our findings and finally in
Sect.~\ref{hd56:sec:summary-conclusions} we give a summary and
conclusions.

\section{Stellar parameters}
\label{hd56:sec:stellarparameters}
\begin{table}
  \caption{UV to IR photometric data of HD~56126.  Units are magnitudes
      unless given explicitly.} 
  \centerline{
    \begin{tabular}{c@{\ } c@{\ } c@{\ } c@{\ } c@{\ } c@{\ } l}
      \hline
      \hline
      \multicolumn{6}{c}{Walraven}\\
      V & V-B & B-U & U-W & B-L \\
      -0.548 & 0.365 & 0.75 & 0.556 & 0.312 & \multicolumn{2}{r}{\citep{1976AnAS...24..413P}}\\
      -0.555 & 0.400 & 0.766 & 0.588 & 0.322 & \multicolumn{2}{r}{(vG 1986)$^a$}\\
      \hline
      \multicolumn{6}{c}{Johnson \citep{1986AnA...155...72V}} \\
      B & V \\
      9.13 & 8.23\\
      \hline
      \multicolumn{6}{c}{Near-IR \citep{1989ApJ...346..265H}} \\
      J & H & K & L & M \\
      6.92 & 6.66 & 6.65 & 6.43 & 6.10\\
      \hline
      \multicolumn{6}{c}{Str\"{o}mgren} \\
      $b$-$y$ & \multicolumn{2}{l}{($v$-$b$)-($b$-$y$)} & \multicolumn{2}{l}{($u$-$v$)-($v$-$b$)} \\
      0.639 & \multicolumn{2}{l}{0.182} & \multicolumn{2}{l}{1.477} & \multicolumn{2}{l}{\citep{1993AnAS..102...89O}}\\
      $u$ & $y$ &\multicolumn{5}{l}{[ergs\,s$^{-1}$cm$^{-2}$\AA$^{-1}$]} \\
      1.03e-12 & 1.71e-12 & & &\multicolumn{3}{r}{\citep{2000ApJ...528..861U}}\\
      \hline
      \multicolumn{6}{c}{Geneva (van Winckel, priv.comm.)} \\
      U   &    B  &     B1  &    B$_2$& V$_1$   &   G \\
      11.010 &  8.337 &  9.452 & 9.616 &  8.970 &  9.213  \\
      \hline
      \hline
      \multicolumn{6}{l}{$^a$\citet{1986AnA...155...72V}}\\
    \end{tabular}}
  \label{hd56:tab:photometry}
\end{table}
HD~56126 is a bright star with a visual magnitude of 8.3. The star has
been classified as a F5 supergiant by \citet{1965LS....C06....0N}.
More recently \citet{1989ApJ...346..265H} published a medium
resolution optical spectrum and determined the spectral type to be
F0-5I, consistent with the previous spectral classification. There are
several photospheric abundance determinations published for HD~56126
\citep{1992AnA...264..159P,1995MNRAS.272..710K,2000AnA...354..135V}.
These studies all demonstrate the low photospheric metallicity of the
star, with a value of [Fe/H] = $-$1.0 and enrichment of C, N, O and
the s{\textendash}process elements. There is less agreement on the
exact amount of enrichment of the CNO-elements. While
\citet{1992AnA...264..159P} and \citet{2000AnA...354..135V} find a C/O
value close to unity, \citet{1995MNRAS.272..710K} determine a lower
abundance of O and derive a C/O value of $\sim$1.5.

The conversion from spectral type to effective temperature
(T$_\mathrm{eff}$) is not clear-cut. Due to the peculiar abundances
one should be critical of applying standard spectral type calibrations
based on solar type abundances. For example, the tables of
\citet{1987AnA...177..217D} quote 6370 for a F5I star with solar
abundances. Values of T$_\mathrm{eff}$ for HD~56126 within a large
range (5900$-$7250~K) are derived by various methods. We give most
weight to the temperatures determined in detailed abundance analyses
using high resolution spectroscopy
\citep{1992AnA...264..159P,2000AnA...354..135V}. These studies yield a
high value of the effective temperature (7000$-$7250~K). This
temperature corresponds to a F1-2I spectral classification, still
within the range derived by \citet{1989ApJ...346..265H}. Such a high
value for T$_\mathrm{eff}$ also fits better with the observed spectral
energy distribution (SED) (see below).

\citet{2000ApJ...544L.141J} have made the case that the low iron
abundance of the star, the high galactic latitude ($b$=$+$10$^\circ$)
and its high radial velocity make it probable that HD~56126 is a
population II star with a zero age main-sequence mass of
$\sim$1.1\,M$_{\odot}$.

The distance ($d$) to HD~56126 is poorly known.  Values between 2 and
3~kpc are used in the literature.  \citet{1999PASJ...51..197Y} derive
a statistical distance of 2.12~kpc from a principle component analysis
based on IRAS photometry and the molecular expansion velocity.
\citet{2000ApJ...534..324K} derive 2.4~kpc from the radial velocity
assuming that the radial velocity is due to the galactic rotation
while \citet{2000ApJ...544L.141J} derive 2.3~kpc from the theoretical
luminosity of 6600~L$_{\odot}$ and the observed flux. In the following
we elected to use $d$=2.4~kpc. Note that this distance yields a
luminosity of 6000~L$_{\odot}$ which is higher than expected
($\sim$3000$-$4000~L$_{\odot}$) \citep[e.g.][]{1988ApJ...328..641B}
for a star with a $\sim$1.1~M$_{\odot}$ ZAMS mass on the basis of the
initial mass-final mass relation and core-mass-luminosity relation.
In view of the uncertain distance, we explicitly give the distance
dependence of the values that we derive.

\section{Observations and data reduction}
\label{hd56:sec:observations}
\subsection{Infrared spectroscopy}
\label{hd56:sec:iso-spectroscopy}
\begin{table}
\caption{Details of the ISO observations used in this study.}
\centerline{
  \begin{tabular}{l@{\ } l@{\ } l@{\ } l@{\ } l}
    \hline
    \hline
    \multicolumn{5}{c}{HD~56126 ISO observations}\\
    %%Column names
    \multicolumn{1}{c}{Instrument}&
    \multicolumn{1}{c}{Obs.$^{a}$}&
    \multicolumn{1}{c}{$\alpha$}&
    \multicolumn{1}{c}{$\delta$}&
    \multicolumn{1}{c}{TDT$^{b}$}
    \\
    %% Column units
    \multicolumn{1}{c}{}&
    \multicolumn{1}{c}{Mode}&
    \multicolumn{1}{c}{(J2000)}&
    \multicolumn{1}{c}{(J2000)}&
    \multicolumn{1}{c}{}
    \\
    \hline
    \hline
    SWS    &  06    &  07\,16\,10.20 &  $+$09\,59\,48.01 & 71802201 \\
    SWS    &  06    &  07\,16\,10.30 &  $+$09\,59\,48.01 & 72201702 \\
    SWS    &  01(3) &  07\,16\,10.20 &  $+$09\,59\,48.01 & 72201901 \\
    LWS    &  01    &  07\,16\,10.20 &  $+$09\,59\,47.80 & 72201802 \\
    \hline
    \hline
  \end{tabular}}
    $^{a}$Observing mode used see \citet{1996AnA...315L..49D} and
    \citet{1996AnA...315L..38C}. Numbers in brackets correspond to the
    scanning speed.\\ 
    $^{b}$TDT number which uniquely identifies each ISO observation.
\label{hd56:tab:obslog}
\end{table}

\begin{figure*}
  \centerline{
    \includegraphics[width=18.0cm]{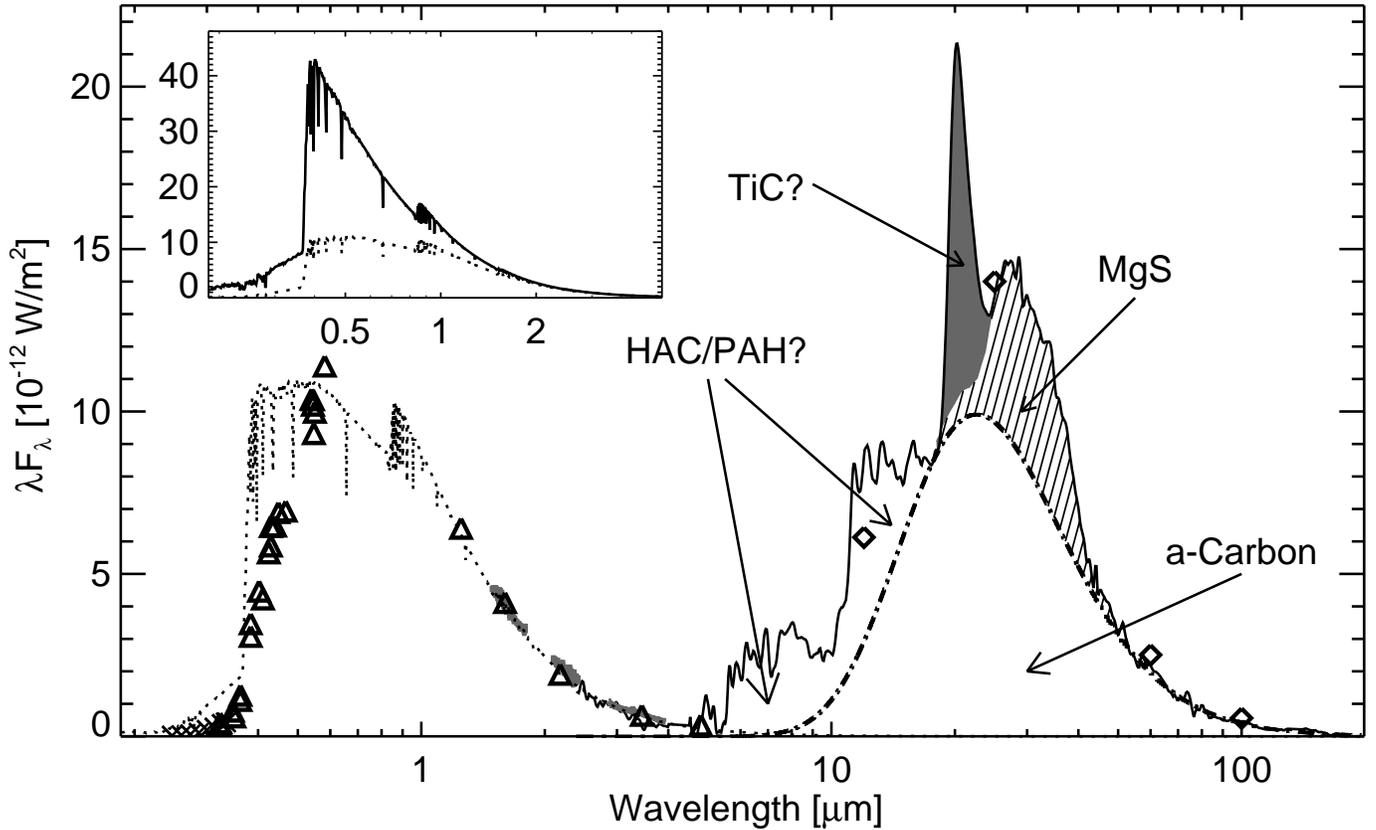}}
  \caption{The spectral energy distribution of HD~56126. We show the
    IUE data (crosses), optical and near-IR photometry (triangles)
    (see Table~\ref{hd56:tab:photometry} for details), IRAS photometry
    (diamonds) and the ISO/SWS and ISO/LWS spectra (solid line). In
    grey we show near-IR spectroscopy from \citet{1990ApJ...360L..23K}
    and \citet{1995AnA...299...69O}. For reference we also show a
    Kurucz model (dotted line) of the central star with
    T$_\mathrm{eff}$=7250~K, $\log g$=1.0 and [M/H] = -1.0
    \citep{1993KurCD...3.....K} reddened with E(B-V)=0.33 using a
    standard extinction law \citep{1979ARAnA..17...73S}. The standard
    extinction law is clearly not applicable. In grey we indicate the
    contribution of the ``21''~$\mu$m feature and the dashed area
    indicates the ``30''~$\mu$m feature due to magnesium sulfide.
    Note that on this scale surface corresponds to energy.  Therefore
    one may readily see that an appreciable amount of energy is
    emitted in the ``21'' and ``30''~$\mu$m features. The inset shows
    the Kurucz model before reddening (solid line) and the reddened
    model atmosphere (dotted line).}
  \label{hd56:fig:sed}
\end{figure*}
For the spectral energy distribution (SED) of HD~56126 we use IR data
obtained with the Short Wavelength Spectrometer (SWS)
\citep{1996AnA...315L..49D} and Long Wavelength Spectrometer (LWS)
\citep{1996AnA...315L..38C} on-board the Infrared Space Observatory
(ISO) \citep{1996AnA...315L..27K}. These data consist of one SWS/AOT01
spectrum from 2.3$-$45~$\mu$m, two SWS/AOT06 spectra covering the
ranges 16.5$-$24 and 23$-$43~$\mu$m and a LWS spectrum from 45 to
200~$\mu$m. Details of the ISO observations are given in
Table~\ref{hd56:tab:obslog}.

The SWS data reduction steps involved the removal of cosmic ray hits.
Next, we combine all available data from the SWS observations in order
to maximise the signal-to-noise ratio.  We correct each separate
observation, subband and detector combination in such a way that its
average equals the average of all available data at the corresponding
wavelength.  Below 12~$\mu$m we apply an offset and above 12~$\mu$m a
scaling to the data. We apply an additional scaling factor of 1.1 to
the data of sub-band 3C (16$-$19.5~$\mu$m) to form a continuous
spectrum. Finally, the data are rebinned to a fixed resolution grid of
$\lambda/\Delta\lambda$=200.

Data reduction of the LWS observation consisted of extensive bad data
removal, rebinning on a fixed resolution grid of $\lambda / \Delta
\lambda$=50 and splicing of the data to form a continuous spectrum.
We apply scaling factors to the data from 45$-$70~$\mu$m and offsets
at longer wavelengths. The data of detector 7 (130$-$140~$\mu$m) are
very bad and we have replaced those by a constant level of 10~Jy, the
average of the much better data on either side of detector 7. The
combined SWS/LWS spectrum together with data from the literature is
presented in Fig.~\ref{hd56:fig:sed}.

\begin{figure}
  \centerline{
  \includegraphics[width=8.8cm]{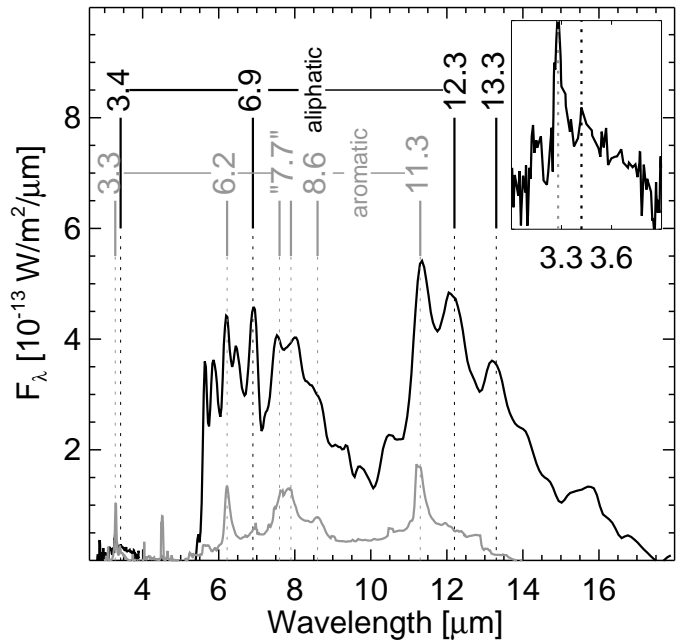}}
  \caption{The 3 to 17~$\mu$m spectrum of HD~56126 after removing the
    contribution of the star and the dust continuum. Several emission
    features due to vibrationally excited C-H and C-C bonds in
    hydrogen-carbon compounds are identified. We have indicated the
    features due to aromatic (grey lines) and aliphatic (black lines)
    structure. The inset shows a blow-up of the 3~$\mu$m region, where
    the dominant aromatic 3.3~$\mu$m and the weaker 3.42~$\mu$m
    signature are observed. For reference we show the spectrum of the
    PN {NGC~7027} as a prototypical PAH spectrum (grey).  The emission
    features of HD~56126 are markedly different from the PAH spectrum
    of NGC~7027.}
  \label{hd56:fig:iso_hac}
\end{figure}
The SED of HD~56126 is double peaked. At wavelengths shorter than
4~$\mu$m we see the stellar photosphere directly. At longer
wavelengths the dust emission dominates. We see continuum emission
which peaks at $\sim$25~$\mu$m with strong broad emission features
perched on top. In Fig.~\ref{hd56:fig:sed} we indicated the broad
continuum due to amorphous carbon (a-Carbon) and the ``21''~$\mu$m,
``30''~$\mu$m and the plateau feature, (tentatively) identified with
titanium carbide, magnesium sulfide, and HAC or PAHs, respectively.

In Fig.~\ref{hd56:fig:iso_hac} we show a detailed view of the plateau
features. We indicate the emission that can be attributed to aromatic
and aliphatic bonds separately. Notice the differences with the
emission spectrum of the planetary nebula NGC~7027 in feature shape
and relative strength of the bands. The aliphatic component is much
stronger in HD~56126 than in NGC~7027.

\subsection{N-band imaging}
\label{hd56:sec:timmi2-imaging}
\begin{figure}
  \includegraphics[width=8.8cm,clip]{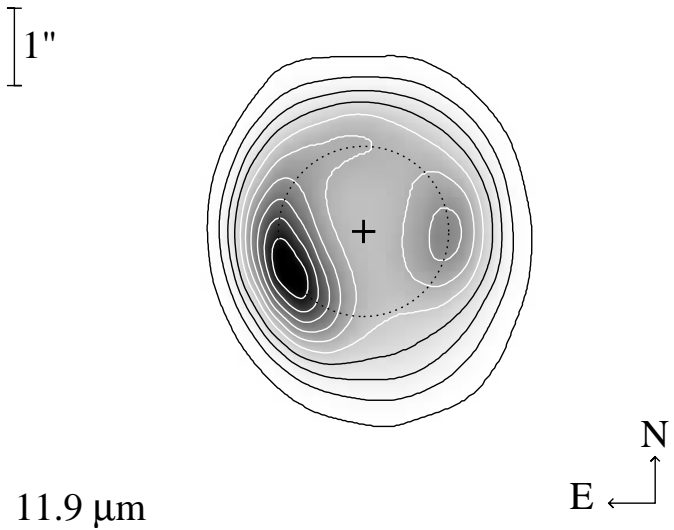}
  \caption{TIMMI2 image of HD~56126 in the 11.9~$\mu$m narrow band
    filter. The scale on the top left indicates 1$^{\prime\prime}$.
    There are three emission maxima on a ring-like structure
    (indicated by a dotted line) at $\sim$1.1$^{\prime\prime}$ from
    the centre (indicated by the '$+$'-sign). The grey-scale shows the
    square of the measured intensity to enhance the contrast of the
    emission maxima. The contours indicate the 5$-$95 per cent
    intensity levels in steps of 10 per cent. The outer contour (5 per
    cent) lies at $\sim$2.4$^{\prime\prime}$ from the centre.}
  \label{hd56:fig:timmi2}
\end{figure}

We also obtained an 11.9~$\mu$m image with the TIMMI2
\citep{timmi2_2000,timmi2_messenger} mid-IR camera attached to the ESO
3.6m telescope at La Silla, Chile on 25 December 2001. The camera is
equipped with a 320x240 pixel array; we applied a pixel scale of
0.3$^{\prime\prime}$/pixel.  We used a chop throw of
15$^{\prime\prime}$ north-south and a nod of 15$^{\prime\prime}$
east-west.  This allowed for both the chopped as well as the nodded
positions to fall onto the detector.  We combined the positive and
negative images resulting from the chopping and nodding positions.
The images were reduced using a shift-and-add technique, where the
shift between individual frames was determined from a least-squares
comparison between the individual frames. The resulting image was then
deconvolved using an empirically determined point spread function
obtained by observing {HD~32887} with the same set-up as HD~56126. We
show the final image in Fig.~\ref{hd56:fig:timmi2}.

The TIMMI2 11.9~$\mu$m image shows a clearly resolved envelope of
$\sim$5$^{\prime\prime}$ diameter. The emission peaks at
1.1$^{\prime\prime}$ to the east from the centre of the nebulae. There
are two other local intensity maxima: in the west and in the north,
although the latter is not clearly resolved. Our 11.9~$\mu$m image is
consistent with other mid-IR imaging studies
\citep{1997ApJ...482..897M,1998ApJ...492..603D,2000ApJ...544L.141J,2002ApJ...573..720K}
that show the same morphology throughout the 8$-$21~$\mu$m wavelength
range. The emission maximum in the north is resolved in the images of
\citet{2002ApJ...573..720K}. We derive the azimuthal averaged
intensity profile from Fig.~\ref{hd56:fig:timmi2} to compare with the
model output.

\section{Basic considerations}
\label{hd56:sec:basics}
Before we describe the details of our model we consider a few basic
properties and the simple constraints that we can derive from them.

\subsection{Energy balance}
\label{hd56:sec:energy-balance}
\begin{figure}
  \centerline{
  \includegraphics[width=8.8cm]{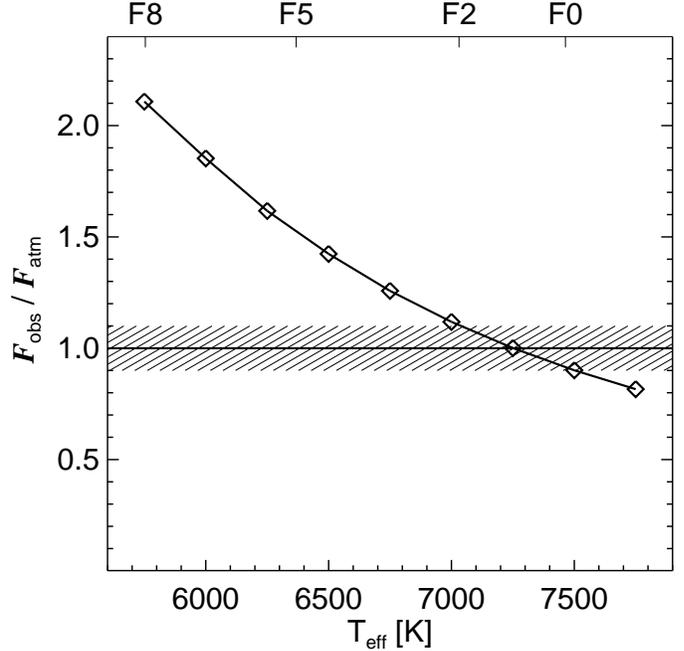}}
  \caption{The ratio of the observed
    integrated flux over the integrated flux from the Kurucz model
    atmosphere scaled to the near-IR observations versus the effective
    temperature of the model atmosphere.  The hatched area indicates
    the range in which this ratio equals unity within the
    uncertainties. We indicate the position of the corresponding
    spectral types taken from \citet{1987AnA...177..217D} for
    supergiants along the top abscissa.}
  \label{hd56:fig:fluxratios}
\end{figure}

In Fig.~\ref{hd56:fig:sed} equal areas under the curve correspond to
equal energy and it is clear that most of the energy is emitted in the
IR.  In Table~\ref{hd56:tab:energybudget} we specify the fractions of
the energy emitted by each of the components in
Fig.~\ref{hd56:fig:sed}. Only 44 per cent of the light comes from the
star directly while 56 per cent has been absorbed in the CSE and
re-emitted in the IR. This implies that the extinction averaged over
all angles around the star must be appreciable. Using a standard
extinction law and a 7250~K Kurucz model atmosphere
\citep{1993KurCD...3.....K} we estimate that the average
A$_\mathrm{V}$ equals 1.1, which corresponds to $\tau_\mathrm{V}$=1.2.
However, such a value for the visual extinction corresponds to hardly
any extinction in the near and mid-IR. Thus the near-IR (2$-$4~$\mu$m)
observations give the true continuum level of the star since no
obvious dust contribution is detected in this wavelength range.
Therefore, for a given effective temperature of the stellar
photosphere there is no freedom in choosing the absolute scale. With
this we can directly compare the integrated flux in the observed SED
to the integrated flux from the model atmosphere. In
Fig.~\ref{hd56:fig:fluxratios} we show the integrated flux in the
\emph{complete} observed SED ($\mathcal{F}_\mathrm{obs}$) divided by
the integrated flux in a Kurucz model atmosphere
($\mathcal{F}_\mathrm{atm}$) as a function of T$_\mathrm{eff}$. As
parameters of the Kurucz models we choose $\log g$=1.0 and a
metallicity of [Fe/H]=-1.0 as determined for this star. For
T$_\mathrm{eff}$ in the range 7000$-$7500~K
$\mathcal{F}_\mathrm{obs}$/$\mathcal{F}_\mathrm{atm}$ is approximately
one. This ratio should be close to unity if the following assumptions
hold. {\it i}) The extinction in the near-IR is negligible. {\it ii})
The extinction along our line of sight is similar to the average value
of the extinction.

\begin{table}
  \caption{The integrated contributions of the different components
    of the spectral energy distribution (see Fig.~\ref{hd56:fig:sed}). We
    list the integrated flux (column 2), the fraction of the total
    luminosity (column 3) and the fraction of the IR luminosity
    (column 4).}
  \centerline{
    \begin{tabular}{l@{\ } D{.}{.}{-1} c c}
    \hline
    \hline
    %%Column names
    \multicolumn{1}{l}{Component}&
    \multicolumn{1}{c}{Flux}&
    \multicolumn{1}{c}{Frac$_\mathrm{total}$}&
    \multicolumn{1}{c}{Frac$_\mathrm{dust}$}
    \\
    %% Column units
    \multicolumn{1}{c}{}&
    \multicolumn{1}{c}{[10$^{-12}$\,W/m$^2$]}&
    \multicolumn{1}{c}{\%}&
    \multicolumn{1}{c}{\%}
    \\
    \hline
            Total    & 33.6 & 100 &  -  \\
        \ \ Star     & 14.8 & 44  &  -  \\
        \ \ Dust     & 18.8 & 56  & 100 \\
    \hline
    \multicolumn{4}{c}{Dust components}\\
    \hline
    \ \ \ \ a-Carbon & 11.6 & 35  & 62  \\
    \ \ \ \ HAC/PAHs & 3.0  &  9  & 16  \\
    \ \ \ \ MgS      & 2.7  &  8  & 14  \\
    \ \ \ \ TiC      & 1.5  &  4  &  8  \\
    \hline
    \hline
  \end{tabular}}
  \label{hd56:tab:energybudget}
\end{table}
The first assumption holds unless there is a large amount of grey
extinction. Extinction caused by particles smaller than the wavelength
at which they absorb falls off steeply when going to longer
wavelengths. This is illustrated by the inset in
Fig.~\ref{hd56:fig:sed} where we show a Kurucz model atmosphere with
T$_\mathrm{eff}$=7250~K before and after reddening with standard
interstellar extinction. The extinction in the UV and visible is
substantial but the flux level at wavelengths beyond 2~$\mu$m is
hardly affected.  Grey extinction is due to dust grains much larger
than the wavelength at which they absorb.  Typically these grains
should have a diameter of at least several $\mu$m in order to cause
wavelength independent extinction up to 4~$\mu$m. It is unlikely that
a large amount of such big grains grow in stellar outflows. Moreover,
the presence of these large dust grains would have strong implications
for the understanding of the complete nebula, most notably the nebular
mass budget. These implications are discussed in
Sect.~\ref{hd56:sec:disc:envel-mass}.

The second condition, concerning the extinction along the line of
sight compared to the average extinction, is more debatable. The
mid-IR images as published by
\citet{1997ApJ...482..897M,1998ApJ...492..603D} and
\citet{2000ApJ...544L.141J} show clear deviations from spherical
symmetry. These asymmetries are due to a non-spherical distribution of
the dust around the star. We can write a simple expression for the
flux observed from the system.
\begin{eqnarray}
  \mathcal{F}_\mathrm{obs} &=& \underbrace{(1-\eta_\mathrm{los}) \,
  \mathcal{F}_\mathrm{atm}}_{\mathrm{UV+visible}}  +
  \underbrace{\eta_\mathrm{aver} \,
  \mathcal{F}_\mathrm{atm}}_{\mathrm{IR}} ~~\Leftrightarrow ~~ \\
  \frac{\mathcal{F}_\mathrm{obs}}{\mathcal{F}_\mathrm{atm}} &=&
  1-\eta_\mathrm{los} + \eta_\mathrm{aver}\,, \nonumber
  \label{hd56:eq:energymodel}
\end{eqnarray}
where $\eta_\mathrm{los}$ is the fraction of the stellar flux absorbed
in the line of sight towards the earth and $\eta_\mathrm{aver}$ is the
fraction of the stellar flux absorbed, averaged over all lines of
sight from the star. In this we assume that the energy absorbed in the
CSE is re-emitted isotropically in the IR.  The energy budget equation
is further constrained by the observational fact that
$\eta_\mathrm{aver}$ = 1.25\,(1-$\eta_\mathrm{los}$). This yields with
Eq.~(\ref{hd56:eq:energymodel})
$\mathcal{F}_\mathrm{obs}$/$\mathcal{F}_\mathrm{atm}$ =
2.25$-$2.25\,$\eta_\mathrm{los}$.

In case $\eta_\mathrm{los}<\eta_\mathrm{aver}$ (i.e. when we are
viewing the star relatively unobscured)
$\mathcal{F}_\mathrm{obs}$/$\mathcal{F}_\mathrm{atm}$ is larger than
unity. This is due to light that the star emits in directions away
from the direction of the earth. When this light is absorbed in the
CSE and partly re-emitted towards earth we measure extra flux.  This
would allow for a lower T$_\mathrm{eff}$ than the 7000~K we derive
from Fig.~\ref{hd56:fig:fluxratios}. However the observations put
strong constraints on the importance of this effect. Mathematically
$\eta_\mathrm{los}$ and $\eta_\mathrm{aver}$ can of course be in the
range 0$-$1, but in practise the UV and visible photometric data
reveal a reddened star and $\eta_\mathrm{los}$ must be larger than
$\tau_\mathrm{V}$ ($\sim$0.33). This is further corroborated by the
fact that 56 per cent of the flux is emitted in the IR part of the
spectrum. In Eq.~(\ref{hd56:eq:energymodel}) a value of
$\eta_\mathrm{los}$ lower than 0.33 requires $\eta_\mathrm{aver}$ to
be over 0.85. This corresponds to the unlikely situation in which the
star is completely surrounded by an opaque CSE with a small hole
through which we happen to view the star unobscured.  Reasonable
values of $\eta_\mathrm{los}$ are within 0.44$-$0.6 which corresponds
to $\mathcal{F}_\mathrm{obs}$/$\mathcal{F}_\mathrm{atm}$ = 1.26$-$0.9
and T$_\mathrm{eff}$=6800$-$7500~K.

It is important to note that both \citet{1997ApJ...482..897M} and
\cite{1998ApJ...492..603D} have constructed models for HD~56126 in
which $\eta_\mathrm{los}>\eta_\mathrm{aver}$ in order to reproduce the
observed mid-IR morphology. This yields a T$_\mathrm{eff}$ in the high
end tail of the range mentioned above. Likewise, interstellar
extinction can be considered as an added contribution to
$\eta_\mathrm{los}$ also yielding a higher effective temperature for
the central star.  \citet[][ chapter 3]{1994Bogaert} has observed
stars in the vicinity of HD~56126 and estimates the interstellar
contribution to the reddening to be E(B-V)$_\mathrm{IS}$=0.04.  We
conclude that a high effective temperature in the range 7000$-$7500~K
fits best with all observed properties of HD~56126. In the following
we use T$_\mathrm{eff}$=7250~K as derived by
\citet{2000AnA...354..135V}.

\subsection{Feature strength}
\label{hd56:sec:feature-strength}
What is also directly apparent from Fig.~\ref{hd56:fig:sed} is the
large fraction of the light emitted in the IR features. We have
indicated the HAC/PAH features and the ``21'' and ``30''~$\mu$m
features.  Together these features carry 40 per cent of the total IR
flux (see also Table~\ref{hd56:tab:energybudget}). Any self-consistent
radiative transfer modelling of the nebula must take absorption and
emission by the feature carriers into account. The reason for this is
twofold.  First, the feature carriers cause an important part of the
reddening of the star. Thus, when comparing model results to the
observed photometry these are essential to assess the validity of the
model.  Second, in the CSE these components absorb a sizeable part of
the stellar light and thus determine, in part, the temperature
structure of the CSE as a whole. The temperature structure of the CSE
is a very important parameter that translates directly into
observables like the mid-IR spectrum and mid-IR surface brightness.

The strength of the IR features as seen in Fig.~\ref{hd56:fig:sed}
also exposes the enigma of this nebula. Several of the observed
features are supposedly carried by materials composed of elements with
low abundances. The key question is: ``Is it possible that trace
species can cause such strong features?''. Note that this question
applies not only to TiC concerning the ``21''~$\mu$m feature but
similarly to the ``30''~$\mu$m feature which is produced by MgS.
While Mg and S are both more abundant than Ti they are relatively
unimportant in comparison to C \citep{2000AnA...354..135V} that causes
the continuum.

\section{Model}
\label{hd56:sec:model}
The purpose of this study is to construct a detailed model of the CSE
of HD~56126.  In our modelling we focus on two key questions:
\begin{itemize}
\item The composition of the CSE, i.e. the amount of the various dust
  components present. As we have already outlined in
  Sect.~\ref{hd56:sec:intro} we observe emission from several
  components for which various carriers have been proposed. However,
  which of these components are present has not been firmly determined
  or the amounts of these carriers in the nebula have not been
  quantified.
\item The location of the dust. The latter is important since there is
  considerable disagreement in the literature regarding the evidence
  for a superwind phase (with mass loss rates possibly exceeding
  10$^{-4}$~M$_{\odot}$/yr).  \citet{1997ApJ...482..897M} and
  \citet{1998ApJ...492..603D} describe the mass-loss history with a
  low mass-loss phase followed by a superwind phase.  Recently,
  \citet{2002ApJ...573..720K} have obtained high angular resolution
  mid-IR images and find no evidence for a large-scale sudden ejection
  of material. On the other hand, \citet{2000Sci...288..313V} have
  proposed that the ``21''~$\mu$m feature is carried by TiC clusters
  and that the formation of these clusters requires an short phase of
  extreme mass loss.
\end{itemize}

Note that the observed mid-IR emission is not spherically symmetric
but shows three intensity maxima along the limb-brightened inner edge
of the dust shell
\citep{1997ApJ...482..897M,1998ApJ...492..603D,2000ApJ...544L.141J,2002ApJ...573..720K}
(see also Fig.~\ref{hd56:fig:timmi2}). We ignore this non-sphericity
in our model because doing so greatly reduces the number of free
parameters without compromising our analysis of the dust composition
and density as a function of distance from the star.  Because the
nebula is largely optically thin, the deviations from spherical
symmetry do not influence our conclusions.  To confine the radial
extent of the dust we compare the model results with the azimuthal
averaged intensity profiles derived from the mid-IR images.

To model the absorption and emission of the circumstellar dust we use
the dust radiative transfer program \textsc{MODUST}.  This code solves
the radiative transfer equation in spherical geometry subject to the
constraint of radiative equilibrium. This yields the temperature of
the dust. The code allows for the presence of several different dust
components of various grain sizes and shapes. The density of the
grains as a function of distance can be prescribed in various ways.
\textsc{MODUST} yields both the model SED and intensity maps. We refer
to \citet{2000AnA...360..213B} and \citet{Bouwman_PhD} for a
description of techniques used in \textsc{MODUST}.  The following
parameters for modelling the HD~56126 system are used:
\begin{itemize}
\item The input spectrum of the central star is given by a line
  blanketed Kurucz model atmosphere \citep{1991sabc.conf..441K}. We
  use T$_\mathrm{eff}$=7250~K and $\log(g)$=1.0. With these parameters
  and the observed total integrated light of 33.6\,10$^{-12}$ W/m$^2$
  the radius and luminosity of the star are related to the distance
  by:
  \begin{equation}
    R_{\star} = 20.5\,d ~~\mathrm{and}~~ L_{\star} = 1051\,d^2,
  \end{equation}
  where $R_{\star}$ is the stellar radius in units of solar radii,
  $L_{\star}$ is the stellar luminosity in units of solar luminosities
  and $d$ is the distance to the earth in kiloparsecs.
\item The location of the dust is constrained by mid-IR imaging. The
  intensity maximum which is observed at $\sim$1.1$^{\prime\prime}$
  from the star corresponds to the inner edge of the dust shell in the
  case of optically thin emission, which is certainly satisfied at IR
  wavelengths.  Taking into account the point-spread-function (PSF) of
  the telescope we find that the inner edge is located at
  $\sim$1.2$^{\prime\prime}$. Because dust emission is clearly
  observed in all mid-IR images out to a $\sim$2.6$^{\prime\prime}$
  radius this constitutes a lower limit for the extent of the dust
  shell. However, the upper limit of the extent is not well
  constrained by current imaging and in the following we will treat
  the outer radius as a free parameter.
\item The dust density ($\rho$) as a function of distance from the
  star ($r$) is specified in the form:
  \begin{equation}
    \rho = \rho_0 \, \left( \frac{r}{r_0} \right)^{-p},
  \end{equation}
  where $\rho_0$ is the dust density at the location of the inner edge
  of the shell ($r_0$) and the power $p$ determines the steepness of
  the density profile. For a time-independent mass loss $p$=2. If the
  mass-loss rate has increased over time $p>$2.
\item The grain size distribution for each separate dust component can
  be specified with a power-law in the form:
  \begin{equation}
    n(a) \propto a^{-q}; ~~ (a_\mathrm{min} \, \le \, a \, \le \,
    a_\mathrm{max}),
  \end{equation}
  where $a$ is the grain radius, $n(a)$ it the number density of
  grains with radius $a$, $q$ is the power and $a_\mathrm{min}$ and
  $a_\mathrm{max}$ are the minimum and maximum grain radius,
  respectively.
\item We choose to keep the dust composition constant as a function of
  distance from the star. This means that the \emph{relative} fraction
  of each grain material component, specified by chemical composition
  and size is fixed throughout the nebula.  The dust shell consists of
  separate grain populations, i.e. each dust grain contains one
  material, implying that separate species can have different
  temperatures at the same location.
\end{itemize}

\section{Optical properties}
\label{hd56:sec:opticalprops}
In this section we discuss the optical properties of the various
solid-state components that we use to model the HD~56126 system. Some
of these materials have been well characterised in the laboratory over
the broad wavelength range required for radiative transfer modelling.
For other materials these data are not available and we describe how
we treated the optical properties of these materials.

\subsection{Carbonaceous compounds}
\label{hd56:sec:carb-comp}
There is general consensus that the mid-IR continuum in these sources
is due to carbon-based dust grains, possibly with inclusion of
hydrogen. The UIR bands at $\sim$3.3, 6$-$9 and 10$-$17~$\mu$m,
indicated in Fig.~\ref{hd56:fig:sed} and \ref{hd56:fig:iso_hac}
correspond to the vibrational resonances of aromatic and aliphatic C-H
and C-C bonds. Despite this general consensus, the exact nature of the
carrier is under much debate. This is partly due to the many different
structures in which carbon can appear. There are three possible
bonding types between the carbon-atoms: sp$^1$, the aromatic sp$^2$
bond, as in graphite, and the aliphatic sp$^3$ bond, as in diamond. It
is due to this diversity of bonding types of carbon and the resulting
diversity in material properties of the various carbonaceous compounds
that carbonaceous dust in space has evaded firm identification. In the
following we consider amorphous carbon and hydrogenated amorphous
carbon.

\subsubsection{Amorphous carbon}
\label{hd56:sec:amorphous-carbon}
The most prominent dust component that we observe is the mid-IR
continuum. It carries $\sim$60 per cent of the IR flux. This emission
is generally attributed to amorphous carbon (a-C). We use the values
of the complex refractive index of amorphous carbon as published by
\citet{1993AnA...279..577P}.  These data range in wavelength from 0.1
to 800~$\mu$m, which is sufficient to properly model the absorption
and emission properties.

\subsubsection{Hydrogenated amorphous carbon}
\label{hd56:sec:hac}
The emission between 3$-$4 and 11$-$17~$\mu$m is indicative of
vibrationally excited C-H bonds. This indicates that the CSE contains
a compound of carbon and hydrogen. When looked at in detail both
emission from H-atoms bound to sp$^2$ and sp$^3$ sites is observed
\citep{1990ApJ...365L..23B,2001ApJ...554L..87K}. Such a mixture of
hydrogen bonds is generally found in hydrogenated amorphous carbon
(HAC). HAC is a likely condensate in H-rich and C-rich circumstellar
ejecta. Therefore we include HAC in our model.

There is extensive literature on the properties of HAC in the material
and surface science literature, where it is commonly referred to as
\emph{a-C:H} or \emph{diamond-like amorphous carbon}. HAC films have
received much attention as optical coatings because they can be
manufactured to be relatively transparent in the 2$-$6~$\mu$m range;
see \citet{Robertson2002}. As a result of this interest, such films
have been studied in great detail: we refer to \citet{Robertson2002}
for an excellent review of HAC properties.  In an astrophysical
context, the properties of HAC have been extensively discussed by
Duley and co-workers
\citep[e.g.][]{1984ApJ...287..694D,1997ApJ...490L.175S,2001ApJ...558L.129G}.
Here we only discuss the optical properties relevant for the current
study. Note that because of its application as optical coating the
material research has been focused on those materials with low IR
absorption.  The more graphite-like materials, with high IR absorption
levels (see below), have received much less attention.

HAC is a semiconductor. Its structure can be represented as sp$^2$
bonded clusters surrounded by sp$^3$ regions.  The optical properties
from the near-UV to the near-IR are determined by the $\pi$-electrons
of the graphitic sp$^2$ bonds. This is due to the fact that the
$\pi$-electrons are weaker bound than the $\sigma$-electrons of the
sp$^3$ hybridisation. By varying the formation conditions, the
sp$^2$/sp$^3$ ratio can be varied, which in turn allows to tune the
absorptivity of HAC \citep{1983ApPhL..42..636D}. It is shown that the
near-UV to visual absorption is determined by the size of the aromatic
domains in the material
\citep[e.g.][]{Angus1986,Robertson1986,Robertson2002}.  HAC contains a
range of sizes of the aromatic domains and the level of absorption
depends on the size of the largest aromatic clusters present in the
structure at the concentration of a few per cent.

In the visual range, HAC absorptivity has an empirically determined
dependence on photon energy and is usually expressed in the form of
the so-called Tauc law \citep{Tauc}:
\begin{equation}
  \alpha(E)\, E = A_\mathrm{T}\left(E-E_\mathrm{T}\right)^2,
  \label{hd56:eq:tauc}
\end{equation}
where $E$ is the photon energy of the incoming light, $\alpha(E)$ is
the absorptivity as a function of energy, $A_\mathrm{T}$ is a constant
that determines the slope and $E_\mathrm{T}$ is the Tauc gap.
$E_\mathrm{T}$ and the optical band-gap (i.e. the energy level above
which photons couple efficiently to the valence electrons) are
directly linked to the average size of the sp$^2$ domains.

At sub-band-gap energies the electronic absorptivity declines
exponentially and is commonly expressed in an Urbach law:
\begin{equation}
  \alpha = \alpha_\mathrm{U} \exp{\frac{E}{E_\mathrm{U}}},
  \label{hd56:eq:urbach}
\end{equation}
where $E_\mathrm{U}$ is the Urbach energy that determines the width of
the tail, while $\alpha_\mathrm{U}$ is a constant that determines the
level. The width of the tail, i.e. $E_\mathrm{U}$, is determined by
the disorder in the material. Hydrogenated amorphous carbon is a
heterogeneous material with a range of bond types and sizes of sp$^2$
bonded domains. As a consequence, the tail is broad. Typical values of
the important parameters in the above equations are:
$A_\mathrm{T}~\simeq$~300~cm$^{-1}$/eV, $E_\mathrm{T}~\simeq$~0$-$2~eV
and $E_\mathrm{U}~\simeq$~200~meV.

In the IR range, the contribution of the electrons to the absorptivity
is in general small, unless one goes to materials with very large
aromatic domains, the limiting case of which is the conductor graphite
that exhibits a very strong electronic continuum in the IR. The IR
absorption of HAC is dominated by vibrational resonances.

We combine the observed optical properties of different HAC materials
from various authors in order to determine the dependence on
structural parameters from the UV to the IR. We use the H/C ratio as
the parameter that determines the material properties. Note, that
similar variations in the sp$^2$/sp$^3$ ratio as described below are
observed in pure amorphous carbon films \citep[e.g.][]{Robertson2002}.
However, here we study the effect of H/C because the astronomical
spectra clearly show (through the features between 10 and 17~$\mu$m)
that there are H-atoms bound to the C-atoms.

\begin{figure}
  \centerline{\includegraphics[width=8.8cm]{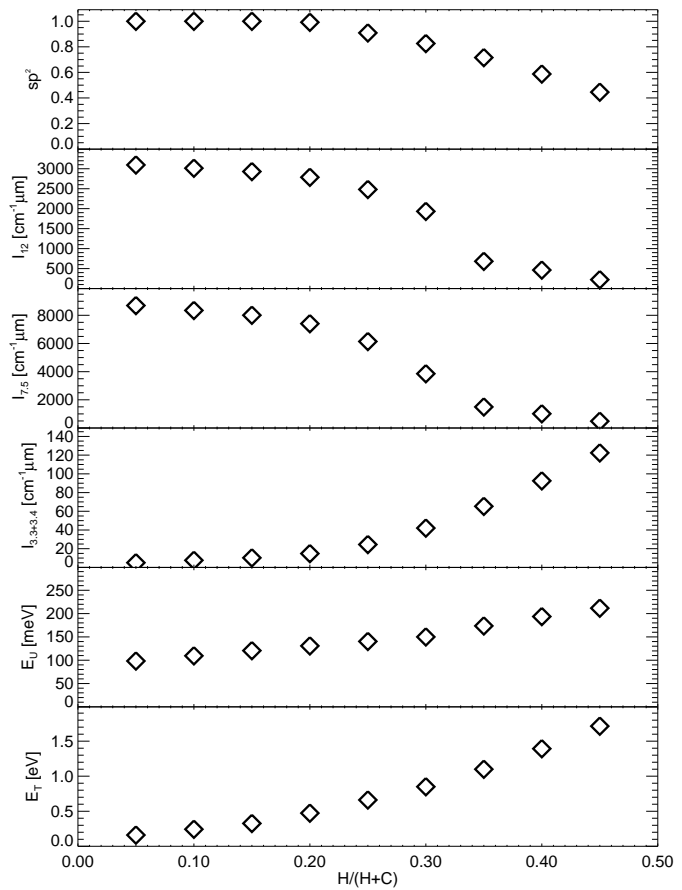}}
  \caption{The optical parameters of HAC as a function of H
    content. The panels show, from top to bottom: 1) the fraction of
    carbon bonded by sp$^2$ bonds. 2,3,4) The integrated strength of
    the 11$-$17, 6$-$9 and the 3.3+3.4~$\mu$m bands, respectively. 5)
    The Urbach energy (see Eq.~(\ref{hd56:eq:urbach})). 6) The Tauc
    energy (see Eq.~(\ref{hd56:eq:tauc})).}
  \label{hd56:fig:hac_parameters}
\end{figure}
\citet{1995PhRvB..5111168C} have studied the optical and IR
absorptivity of HAC films as a function of hydrogen content. These
authors list the parameters for both the Tauc law in the visible and
the Urbach tail dominating the electronic continuum in the IR. For the
HAC materials concerned here, hydrogen incorporation determines the
structural parameters because hydrogen inclusion favours sp$^3$
bonding. Hydrogen incorporation therefore limits the size of the
graphitic domains. Thus, with increasing hydrogen content the
electronic absorption decreases and the optical gap increases.  This
can be seen in Fig.~\ref{hd56:fig:hac_parameters}, where we give the
optical parameters of HAC as a function of hydrogen content.

A clear correlation is found between the optical gap and the strength
of the integrated aromatic plus aliphatic CH stretching feature around
3.3~$\mu$m \citep{1987Dischler}. We use this relation to link the
strength of the CH stretching mode to the electronic continuum level
(cf., Fig.~\ref{hd56:fig:hac_parameters}).

\citet{1995PhRvB..51.9597B} discuss the influence of annealing on the
IR vibrational spectrum of HAC. Annealing removes H and as a result
produces more and more graphite-like HAC. We use the strengths of the
vibrational bands at 3.3, 3.4, 6$-$9 and 10$-$17~$\mu$m for their
series I material to empirically determine the strength of the bands
as a function of H/C ratio.

\begin{figure}
  \centerline{ \includegraphics[width=8.8cm]{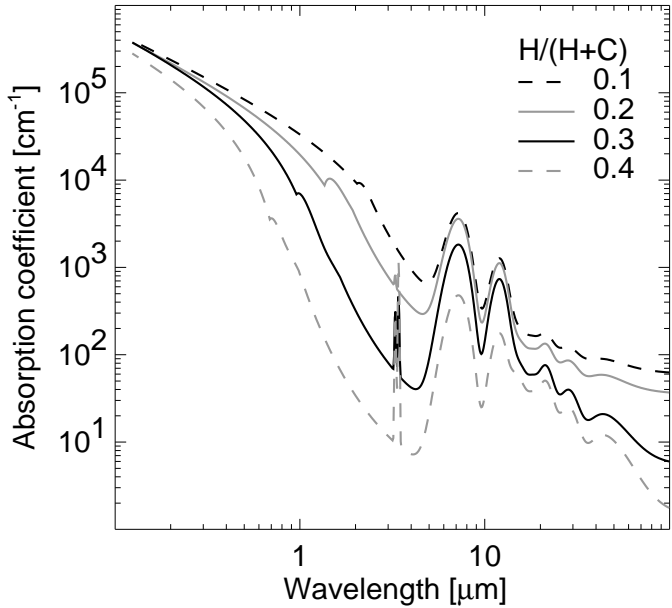}}
  \caption{Absorption coefficient ($\alpha$) of HAC as a function of
    wavelength for changing values of hydrogen content. For details
    concerning the method used to derive the absorptivity, see text.}
  \label{hd56:fig:hac_absorptivity}
\end{figure}
HAC material also exhibits a far-IR continuum due to vibrations.
Unfortunately, these contributions have not been studied well because
they contain little diagnostic value for the material properties.
Following \citet{1993ApJ...415..397S}, we estimate the level of the
far-IR vibrational continuum from the vibrational resonances observed
in aromatic molecules.  \citet{1996AnA...310..297M} have measured the
far-IR spectra of a variety of different PAHs. We use the resonances
these authors find to define an average far-IR continuum.  We apply
the following procedure to construct this average. We take the
wavelength and absorption cross-section per C-atom for each of the 125
resonances listed in Tab.~3 of \citet{1996AnA...310..297M} and
convolve these with a Gauss function with a full-width-at-half-maximum
of 80 cm$^{-1}$.  After adding the contributions we renormalise by
dividing by the total number of molecules (40) in the database and
converting to absorptivity values in cm$^{-1}$ using a typical value
for the density of 2 g/cm$^{3}$.  The far-IR contribution that we
obtain in this way is not completely smooth but shows some structure
(see the 20$-$50~$\mu$m range in
Fig.~\ref{hd56:fig:hac_absorptivity}).  This spectral structure is
most likely due to the fact that we included a biased sample of 'only'
40 molecules in a limited size range to construct the average
absorption.  A larger number of molecules with a wider range of
molecular structures will probably fill in the gaps and yield a
smoother far-IR continuum.  Since this spectral structure is weak
compared to the near- and mid-IR features and is lost in the emission
from other dust components in the circumstellar spectrum, we have
refrained from (arbitrarily) smoothing it out. The optical gap and the
sp$^2$ fraction are correlated as explained above.  The measured
relation for a large number of HAC and a-C materials is given in
\citet[][ Fig.~58]{Robertson2002} from which we derive the sp$^2$
fraction (cf., Fig.~\ref{hd56:fig:hac_parameters}).  The final
estimate for the far-IR vibrational continuum is the product of the
sp$^2$ fraction and the average far-IR ``continuum'' as measured from
PAHs. The estimate that we obtain is a lower limit to the real
vibrational contribution to the far-IR absorptivity because we only
consider the contributions due to the graphitic (sp$^2$) fraction in
HAC, while in reality the sp$^3$ skeleton will also contribute.

We show the dependence of the various parameters on the hydrogen
incorporation in Fig.~\ref{hd56:fig:hac_parameters}. The strong
increase in the Tauc and Urbach energy with H/(H+C) is apparent.
Likewise, the 3.3 and 3.4~$\mu$m CH stretch modes increase with
H/(H+C) fraction.  Notice how the strength of the C-H out-of-plane
bending modes between 11 and 17~$\mu$m (in
Fig.~\ref{hd56:fig:hac_parameters} represented by I$_{12}$) increases
with decreasing H content. This seemingly contradictory behaviour is
caused by the fact that in absolute terms the H bonded to sp$^2$ sites
increases during annealing. In this process some of the H bonded to
the sp$^3$ sites is converted into H bonded to sp$^2$ sites. The
absorptivity per sp$^2$ C-H bond is much larger than per sp$^3$ C-H
and the result is an increase of I$_{12}$. In contrast the aliphatic
CH stretching modes around 3.4~$\mu$m are stronger than the aromatic
modes (3.3~$\mu$m)

The wavelength dependent absorption cross-sections are shown in
Fig.~\ref{hd56:fig:hac_absorptivity}. The main effect is that the
\emph{complete} continuum level from the UV to the IR decreases with
increasing hydrogen content. Because the UV/visible part of the
spectrum decreases less than the IR, HAC with a higher H concentration
will be warmer. While the continuum increases and the CH stretching
modes at 3.3 strengthen with increasing H content, the mid-IR
resonances between 6$-$9 and 10$-$17~$\mu$m diminish in relative
strength. In the following we refer to HAC with H/(H+C) = $x$ as
HAC$_x$.

\subsection{Coal}
\label{hd56:sec:coal}
\citet{1996ApJ...464..810G} have carried out a spectroscopic
comparison between the emission features of carbon-rich post-AGBs and
terrestrial coals . This comparison looks very promising. Of course,
the CSE of post-AGB objects do not contain coals as such since coals
are the product of the biological cycle in the earth-ecosystem.
However, it may be worthwhile to consider materials with a very
different formation mechanism as analogues to circumstellar materials.

In stark contrast to HAC, the optical properties of coal have been
studied much less and no spectra that cover the full wavelength range
from the UV to the IR for one single sample are available. Moreover
there is a large spread in the derived complex refractive index of
coals; see for example the comparisons made in \citet{Manickavasagam}
and \citet{1996Bhattacharya}. This large spread is probably due to the
wide variety of coal structural properties, even for materials that
are labelled the same. When studying more controlled and limited
samples of coal similar trends as described in
Sect.~\ref{hd56:sec:hac} are recovered.  For example,
\citet{1996Ibarra} find a very similar dependence of the aromatic
fraction on the hydrogen content as found in HAC. Likewise, similar
behaviour for the strength of the 3.3 versus the 3.4~$\mu$m band as a
function of aromaticity (towards anthracite) is found
\citep{1996Cerny}.  \citet{1984Brewster} find a significant increase
in the free electron absorptivity upon graphitisation, again the same
as is found in HAC.  Unfortunately, there are no coal data available
that cover the UV to the IR obtained using one single sample. However,
from the similar behaviour found in coals and in HAC we conclude that
our HAC parametrisation applies to coals as well.

One important difference between coals and HAC is the inclusion of
atom impurities, primarily O and mineral phases in coals. Of course,
our HAC model will not be able to reproduce the effects of these
impurities. The effects of impurities in the CSE of carbon stars must
however be limited because impurities have clear spectroscopic
signatures like strong bands due to O-H ($\sim$3~$\mu$m) and C=O
groups (5.8~$\mu$m) \citep{1998Puente}. Such signatures are not
detected in the astronomical spectrum. We conclude that in so far coal
can be used as an analogue for astronomical carbonaceous grains, its
spectral properties can be well represented by those of HAC.

\begin{figure}
  \centerline{\includegraphics[width=8.8cm]{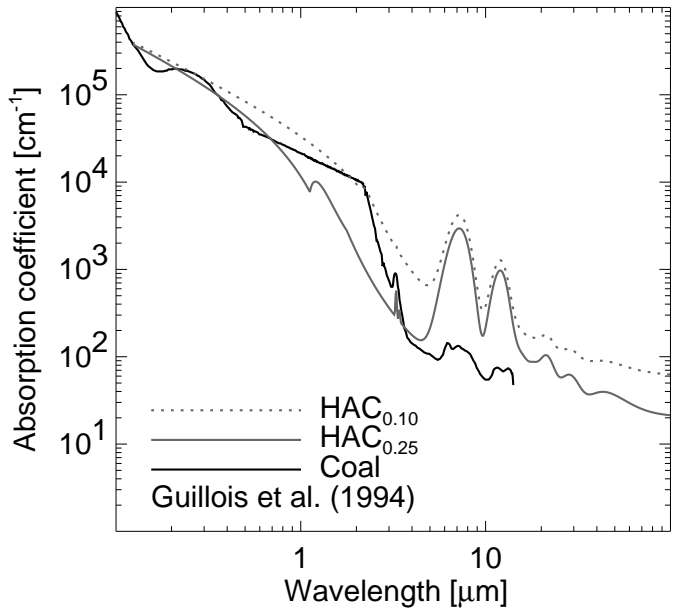}}
  \caption{A comparison between the optical properties of HAC as
    derived from our model and ``standard coal'' as taken from
    \citet{1994AnA...285.1003G}. The continuum of coal lies between
    the H/(H+C)=0.10 and the H/(H+C)=0.25 curves.}
  \label{hd56:fig:alphas}
\end{figure}
In Fig.~\ref{hd56:fig:alphas} we show a comparison between the optical
properties as derived from our model and the optical properties of
``standard coal'' taken from \citet{1994AnA...285.1003G}. Their coal
optical properties are not derived from one single sample and the data
from 0.7 to 3~$\mu$m are not measured but interpolated. The
interpolated optical properties in coal are higher than the Urbach
tail in HAC$_{0.25}$. At the same time, the IR absorption properties
adopted by \citet{1994AnA...285.1003G} are much weaker than the
HAC$_{0.25}$ features. This difference will give rise to a higher
temperature for the ``coal'' than for HAC.

\subsection{Magnesium sulfide}
\label{hd56:sec:mgs}
\citet{1994ApJ...423L..71B} have published the measured values of the
complex refractive index ($n$ and $k$ values) in the 10$-$500~$\mu$m
range of Mg$_x$Fe$_{(1-x)}$S with $x$=0.0, 0.1, 0.5, 0.75, 0.9. They
find little difference between the samples with $x$=0.75 and $x$=0.9
and these authors suggest that the $x$=0.9 sample is a good substitute
for MgS. In order to reliably model the MgS contribution we need the
optical properties in the UV to near-IR range as well.  These data are
unfortunately not available.  We have opted to assume a grain size
($a_\mathrm{model}$) and calculate the mid-IR absorption cross section
from the $n$ and $k$ values in a continuous distribution of ellipsoids
(CDE) shape distribution following \citet[][ Chapter
12]{BohrenHuffman}. We use the CDE distribution because this yields an
emission profile closest to the observed profile in many carbon-rich
evolved stars \citep[e.g.][]{1994ApJ...423L..71B,1999AnA...345L..39S},
see \citet{2002A&A...390..533H} for an elaborate discussion. By
dividing the obtained cross section by the geometrical cross section
we obtain the dimensionless absorption efficiency $Q(\lambda)$. In the
wavelength range 0 to 1~$\mu$m we assume Q=1, from 1 to 2~$\mu$m we
decrease Q linearly to 0 and from 2 to 10~$\mu$m we take Q=0.

The assumptions above are of course unrealistic and will (most likely)
overestimate the visual absorption for a given grain size. However,
this does not influence the derived mass as we can easily derive the
MgS grain temperature from the emission profile
\citep[See][]{2002A&A...390..533H}. In thermal balance the MgS
temperature is determined by the ratio of visual over the IR
absorption efficiency.  In general the IR absorption efficiency
decreases more rapidly with decreasing grain size than the UV and
visible absorption efficiency.  Therefore, using the MgS temperature
derived from the peak of the feature, there exists a grain radius
($a_\mathrm{CSE}$) for which the absorption equals the observed
emission in the MgS band.  This value of $a_\mathrm{CSE}$ will be
smaller or equal to the initial assumed grain size $a_\mathrm{model}$.
The mass that we derive does not depend on the grain size since the
individual grains are optically thin at 30~$\mu$m and therefore the
emission scales with the mass (see also
Sect.~\ref{hd56:sec:mgs-mass}).

\subsection{Titanium carbide nano-crystals}
\label{hd56:sec:tic}
The identification of TiC as the carrier of the ``21''~$\mu$m feature
is based on the wavelength coincidence of a strong resonance found in
TiC nano-crystals with the ``21''~$\mu$m feature
\citep{2000Sci...288..313V}. The optical properties of such
nano-crystals have so far not been determined in the laboratory.  The
optical properties of TiC in the bulk have been studied extensively
\citep{1966PhRv..147..622L, 1975PhRvB..12.1105A, 1975PhRvB..12.5465I,
  1980PhRvB..22.3991L, 1984PhRvB..30.1155P, 1990PhRvB..42.4979K,
  1993koide, 1996PhRvB..54.1673D, 2002espinosa}.  Although these
studies are focused on the high energy range (the longest wavelength
presented is 12~$\mu$m), they consistently find a large free electron
contribution to the optical properties in the infrared.
\citet{Henning_TiC2001} present the first mid-IR spectroscopy of bulk
TiC. Their spectrum indeed shows a strong mid-IR continuum due to free
electrons. They find no resonance at 20.1~$\mu$m and only a very weak
resonance near 19~$\mu$m, much weaker than the resonance found in
nano-TiC.

In principle one can predict how some changes in spectral behaviour
occur when comparing bulk optical properties with those of small
($<$200 atoms) clusters \citep{Kreibig1995}. One of the effects is the
limitation on the electron mean free-path \citep{1969Kreibig}.  This
effect is caused by ``free'' electrons that are scattered off the
surface of the nano-crystal with a frequency that exceeds the
frequency of collisions in the bulk material. In order to
quantitatively calculate changes in the optical constants with respect
to the bulk optical properties requires separating the contribution of
non-conducting electrons to the optical absorption from the
free-electron contribution. For example, a detailed description of how
to apply this correction for silver is given in \citet{1997Alvarez}.
This separation is based on fitting the optical properties over a
wavelength range where the free-electron contribution to the
absorption provides the dominant source of opacity.  However, in TiC
there is no wavelength range were this is clearly the case. For
example \citet{1993koide} find (besides free-electron opacity)
inter-band transitions, due to non-conducting electrons, up to
12~$\mu$m; the longest wavelength they measured. As a consequence, the
estimated contribution due to free electrons derived in one wavelength
range yields results that clearly over- or underestimate the
free-electron contribution in other wavelength ranges.

Another observed effect is the strengthening of the 20~$\mu$m
resonance. If we translate the strength of the 20~$\mu$m feature in
nano-crystalline TiC to bulk samples this resonance would stand out
clearly in the spectrum. Instead, a feature is barely detected
\citep{Henning_TiC2001}.  This immediately shows that bulk TiC cannot
explain the observed ``21''~$\mu$m feature. \emph{Therefore, if TiC is
  the carrier of the ``21''~$\mu$m feature the TiC must be in grains
  or in polycrystalline domains smaller than a few hundred atoms.}

Other quantum effects may also be important in nano-crystalline
materials. The localisation of the electrons may result in a material
that behaves like a semi-conductor or insulator in the
nano-crystalline phase even if the bulk material is a good conductor.
Such behaviour has been observed for example in mercury
\citep{1998PhRvL..81.3836B}.

The issues outlined above leave it impossible to derive reliable
optical constants for TiC nano-crystals in a simple semi-empirical
way.  It is clear that these issues can only be satisfactorily
resolved by laboratory measurements of the optical properties of
nano-TiC. Still, for our radiative transfer calculations some
assumptions about the optical properties of TiC must be made.  For the
emission of the TiC grains we have adopted an approach similar to the
one outlined above for MgS. We use the absorption profile and strength
for TiC nano-crystals as presented by \citet{2000Sci...288..313V} and
assume a constant level for the UV/visible absorption efficiency. This
yields a TiC mass as a function of the UV/visible absorption level.

\section{Radiative transfer modelling}
\label{hd56:sec:model-results}
\subsection{Initial model}
\label{hd56:sec:initial-model}
We first model only the dust continuum due to amorphous carbon. We use
the parameters as described above. As a starting point for the fitting
procedure we use a constant mass loss and outflow velocity profile
($\rho \propto r^{-2}$) during a period of 10\,000 years. This
corresponds to an envelope with radius of $\sim$10$^{\prime\prime}$.
We use a single grain size of 0.1~$\mu$m. Amorphous carbon grains of
this size reach the highest temperature in radiative equilibrium,
given the radiation field of the central star. We vary the density at
the inner edge of the envelope ($\rho_0$) to fit the emission at the
shortest wavelengths of the SWS spectrum. The emission at the shortest
wavelengths is sensitive only to the warmest dust which allows us to
derive the mass of the warmest amorphous carbon grains. We choose
9$-$10 and 18~$\mu$m as reference points. Below 9~$\mu$m the data are
noisy and/or the contributions from the star and the features are more
difficult to estimate while at longer wavelengths the emission is
partly due to cooler grains.

\subsection{Cold component/extent of the nebula}
\label{hd56:sec:cold-comp-nebula}
The model that we construct in this way over predicts the flux levels
beyond 40~$\mu$m by $\sim$10$-$20 per cent, i.e. the model contains
too much cold material. At the same time the modelled 11.9~$\mu$m and
18.2~$\mu$m spatial intensity profiles drop more steeply than observed
within the closest 2.6$^{\prime\prime}$.  The same has been noted at
shorter wavelengths by \citet{1997ApJ...482..897M} and
\citet{2000ApJ...544L.141J}.  \citet{1997ApJ...482..897M} use $\rho
\propto r^{-1}$ ($p$=1) in order to fit their mid-IR images.  However,
this introduces even more cold dust located far from the star. We
compare the observed 18.2~$\mu$m intensity profile and the model
profiles of the $p$=1,2 models in Fig.~\ref{hd56:fig:cbf_intensity},
below. The $p$=2 model peaks too much to the centre of the nebula and
produces too little flux in the outer nebula, while the $p$=1 model is
much closer to the observed intensity profile. Because the
`$p$=2'-model over predicts the flux levels at the longest wavelengths
together with the fact that the observed intensity profile from the
inner envelope (1$-$2.6$^{\prime\prime}$) requires $p$=1, prompts us
to limit the extent of the nebula.

From an observational point of view there is no evidence that the
envelope extends beyond $\sim$2.6$^{\prime\prime}$. All available
images show only emission from this inner region. Note, that this is
not conclusive evidence because the available mid-IR images are all at
wavelength shorter than 25~$\mu$m, i.e. at the Wien side of the dust
emission (see Fig.~\ref{hd56:fig:sed}). This emission is extremely
sensitive to the dust temperature and the cool dust far away from the
star does not contribute much. Deep optical images that show scattered
light from the nebula \citep{2000ApJ...528..861U} show no evidence for
emission outside the $\sim$3$^{\prime\prime}$ radius.  Noteworthy in
this respect is that the very high quality continuum images at 10 and
18~$\mu$m, as presented by \citet{2002ApJ...573..720K}, are very
similar in morphology and \emph{size} while a dust envelope that
continues beyond 3$^{\prime\prime}$ would cause the 18~$\mu$m images
to be $\sim$25 per cent larger than the 10~$\mu$m image, where the
size is measured at 10 per cent of the peak intensity. This `growing'
with increasing wavelength is not observed.

We construct a model with the same parameters as above but with $p$=1.
This model fits the observed 11.9 and 18.2~$\mu$m intensity profiles
much better than a $r^{-2}$ law. From this we derive an upper limit on
the outer radius of the envelope of 4$^{\prime\prime}$. A larger
envelope produces too much far-IR flux.

Up to now we only considered 0.1~$\mu$m amorphous carbon grains. In
reality grains of different sizes are expected to form
\citep{1989AnA...223..227D}. We use a standard MRN size distribution
\citep{1977ApJ...217..425M}: $n(a) \propto a^{-3.5}$. We use
0.01~$\mu$m and 1.0~$\mu$m as the smallest and largest grain radius,
respectively.

As described above the outer radius of the envelope is determined
spectroscopically from the far-IR flux levels, i.e. the relative
amounts of warm and cool a-C. A grain-size distribution affects these
relative amounts because grains of different size reach different
temperatures. Therefore, by using a grain-size distribution the outer
radius can be further constrained.  With the grain-size distribution
as described we find that with a size of 4$^{\prime\prime}$ the
predicted far-IR fluxes are 50 per cent too high. We find that the
best fitting outer radius of the envelope is 2.6$^{\prime\prime}$.  We
use this outer radius in the following analysis.

\subsection{HAC temperature}
\label{hd56:sec:hac-temperature}
\begin{table}
  \caption{HAC temperature and band strength as a function of
    H/(H+C). H/C is the relative number of H-atoms to
    C-atoms. T$_\mathrm{max}$ is the maximum temperature of the
    grains. F$_\mathrm{6-17,HAC}$/F$_\mathrm{6-17,obs}$ is ratio of
    the integrated flux emitted in the plateau features in the model
    over the observed integrated flux in the plateau features,
    assuming equal mass of HAC$_{X}$ and amorphous carbon.} 
  \centerline{
    \begin{tabular}{l@{\hspace{1cm}} D{.}{.}{-1} r D{.}{.}{-1}}
      \hline
      \hline
      %%Column names
      \multicolumn{1}{c}{}& 
      \multicolumn{1}{c}{H/C}&
      \multicolumn{1}{c}{T$_\mathrm{max}$}& 
      \multicolumn{1}{c}{F$_\mathrm{6-17,HAC}$/F$_\mathrm{6-17,obs}$}
      \\
      %% Column units
      &
      &
      \multicolumn{1}{c}{[K]}&
      \\
      \hline
      a-C          & 0    & 170 & $-$ \\
      HAC$_{0.10}$ & 0.11 & 188 & 1.4 \\
      HAC$_{0.20}$ & 0.25 & 190 & 1.7 \\
      HAC$_{0.30}$ & 0.43 & 195 & 1.4 \\
      HAC$_{0.35}$ & 0.54 & 205 & 0.8 \\
      HAC$_{0.40}$ & 0.67 & 205 & 0.5 \\
      HAC$_{0.45}$ & 0.82 & 205 & 0.2 \\
      \hline
      \hline
    \end{tabular}}
  \label{hd56:tab:hac_temperature}
\end{table}
We include HAC as an additional component to a-C. We vary H/C and the
amount of HAC to best fit the observations. We do not aim to fit the
features in detail but aim to determine if HAC reaches the temperature
needed to explain the relative band strength of the broad emission
features. With increasing H/C both the UV/visible and the IR
absorptivity decrease. Because the IR absorptivity decreases more
steeply, grains with a high H/C will be warmest. On the other hand,
the amount of HAC required to explain the energy emitted in the
features increases with H/C content because HAC with high H/C absorbs
the stellar light less efficiently. In
Table~\ref{hd56:tab:hac_temperature} we give the temperature that HAC
grains reach at the inside of the envelope and the strength of the
features as a function of hydrogen content. All HAC grains are warmer
than a-C grains. In the H/(H+C)=0.35$-$0.45 range HAC grain reach a
maximum temperature of $\sim$205~K.
F$_\mathrm{6-17,HAC}$/F$_\mathrm{6-17,obs}$ gives the integrated
strength of the plateau features relative to the observed strength
assuming that the total HAC mass is equal to the a-C mass.  Note, that
we list the \emph{total} flux emitted in the two plateau features. In
all cases the flux emitted in the 10$-$17~$\mu$m band dominates and
the 6$-$9~$\mu$m band is too weak compared to the observations, i.e.
the temperature is too low.

As can be seen from Table~\ref{hd56:tab:hac_temperature}, HAC$_{0.35}$
is optimal in the sense that it reaches a high temperature while
requiring `only' equal mass as contained in a-C. To get the band
strength of the 6$-$9 and 10$-$17~$\mu$m features right we would have
to decrease the inner radius of the envelope. The band strengths are
correct only when we would decrease the inner radius to
$\sim$0.5$^{\prime\prime}$. At 0.5$^{\prime\prime}$ the HAC grains
reach a temperature of $\sim$290~K.  However, such a small inner
radius of the dust envelope is inconsistent with the available mid-IR
images.

The comparison with the observations becomes worse in the 3~$\mu$m
region. In our model HAC$_{0.35}$ produces no discernible aromatic
3.28~$\mu$m or aliphatic 3.42~$\mu$m feature while both are observed
\citep{1990ApJ...360L..23K}. We conclude that HAC grains in radiative
equilibrium do not reach the temperature required to explain the
observed strength of the 3 and 6$-$9~$\mu$m features relative to the
10$-$17~$\mu$m plateau features.

\subsection{Magnesium sulfide}
\label{hd56:sec:mgs-mass}
Before we include MgS in our model we derive the mass of the MgS
component in a simple, relatively model independent, way. In the limit
of grains much smaller than the wavelength (Rayleigh limit), the
absorptivity of a grain scales linearly with the mass of the grain. In
the optically thin case the observed flux can thus be expressed in
terms of distance from the earth ($d$), grain temperature ($T$) and
total mass ($M$) alone.  This can easily be derived in the following
way.  Consider the flux density emitted by a single grain ($f_{1}$) of
radius $a$ and temperature $T$:
\begin{eqnarray}
  f_{1}(\lambda,a,T(a)) &=& 4 \pi a^2 Q(\lambda,a) \, \pi
                     B(\lambda,T(a)) \\
                     &=&3V(a) \frac{Q(\lambda,a)}{a} \, \pi
                     B(\lambda,T(a)),\nonumber 
\end{eqnarray}
where $\lambda$ is the wavelength, $Q(\lambda,a)$ is the absorption
efficiency; thus $4 \pi a^2 Q(\lambda,a)$ is the effective radiating
surface of the grain, $\pi B(\lambda,T(a))$ is the black body flux
density emitted per unit area and $V(a)$ is the volume of the grain.
In the Rayleigh limit $\frac{Q(\lambda,a)}{a}$ is independent of $a$
and therefore $T$ is independent of $a$ and we can write:
\begin{eqnarray}
  f_{1}(\lambda,a,T) &=& 3 \frac{M(a)}{\rho} C(\lambda) \, \pi
  B(\lambda,T) \Rightarrow \\
  f_\mathrm{tot}(\lambda,T) &=& \sum_i 3 \frac{M(a_i)}{\rho} C(\lambda) \, \pi
  B(\lambda,T) \nonumber \\
  &=& 3 \frac{M_\mathrm{tot}}{\rho} C(\lambda) \, \pi 
  B(\lambda,T),\nonumber
\end{eqnarray}
where $M(a)$ is the mass of the grain, $\rho$ is the density of the
material $C(\lambda) = \frac{Q(\lambda,a)}{a}$ and $f_\mathrm{tot}$ is
the total flux density emitted by all grains, the summation is done
over all grains ($i$) and $M_\mathrm{tot}$ is the total mass of the
grains.  Integrating over wavelength and correcting for the distance
we finally obtain:
\begin{eqnarray}
  F_{\lambda_{0,1}} &=& \frac{1}{4 \pi d^2} \int_{\lambda_0}^{\lambda_1} 3
  \frac{M_\mathrm{tot}}{\rho} C(\lambda) \, \pi B(\lambda,T) \, d\lambda
  ~~\mathrm{and} \\
  M_\mathrm{tot} &=& \frac{4}{3} \, F_{\lambda_{0,1}} \, d^2 \, \rho \, \left(
  \int_{\lambda_0}^{\lambda_1} C(\lambda) \, B(\lambda,T) \, d\lambda
  \right)^{-1}, \nonumber 
  \label{hd56:eqn:Mtot}
\end{eqnarray}
where $F_{\lambda_{0,1}}$ is the observed flux integrated between
$\lambda$=$\lambda_0$ and $\lambda$=$\lambda_1$. The last term in
Eq.~(\ref{hd56:eqn:Mtot}) depends only on $T$ for a given material.
The peak position of the ``30''~$\mu$m feature is a good indicator of
$T$ for MgS \citep{2002A&A...390..533H} and we derive $T_\mathrm{MgS}
\simeq$150~K from the observed peak position in the spectrum of
HD~56126.  Using the measured values $\rho=3 g/cm^3$, $F_{10,50}=2.7
\, 10^{-12}$ W/m$^2$ and $d$=2400~pc we find $M_\mathrm{MgS}=9 \,
10^{-6} \, M_{\odot}$.

If we assume that all S is condensed in the form of MgS we can
calculate the total corresponding envelope mass ($M_\mathrm{env}$).
\citet{2000AnA...354..135V} have measured the atmospheric abundance by
number of both Mg/H and S/H to be $\sim$4\,10$^{-6}$, which yields a
$M_\mathrm{env}$ of 0.06~$M_{\odot}$. The mass of the envelope is
variously estimated by fitting the SED and mid-IR imaging at
0.54~$M_{\odot}$ \citep{1997ApJ...482..897M}, 0.26~$M_{\odot}$
\citep{1998ApJ...492..603D} and 0.4~$M_{\odot}$ that we derive below.
We conclude that the abundances of Mg and S are consistent with the
strength of the ``30''~$\mu$m feature.

Of course, these values of $M_\mathrm{MgS}$ and $M_\mathrm{env}$ are
lower limits.  First, because we have assumed that \emph{all} the MgS
is at 150~K. In reality there will be a range of temperatures. The
colder grains are more ``hidden'' and this increases $M_\mathrm{MgS}$,
see Eq.~(\ref{hd56:eqn:Mtot}).  However, while such a temperature is
too low to be compatible with the observed peak position of the
``30''~$\mu$m feature, decreasing $T_\mathrm{MgS}$ to 100~K increases
$M_\mathrm{MgS}$ only by a factor 5, which still yields a reasonable
value for $M_\mathrm{env}$.  Second, because we assume that all Mg-
and S-atoms are condensed into MgS to derive $M_\mathrm{env}$ from
$M_\mathrm{MgS}$.  We point out that there are no other Mg or S
bearing components identified in the nebula. We conclude that from a
simple analysis the strength of the MgS feature as observed in
HD~56126 is compatible with the available Mg- and S-atoms.

\begin{figure}
  \centerline{\includegraphics[width=8.8cm]{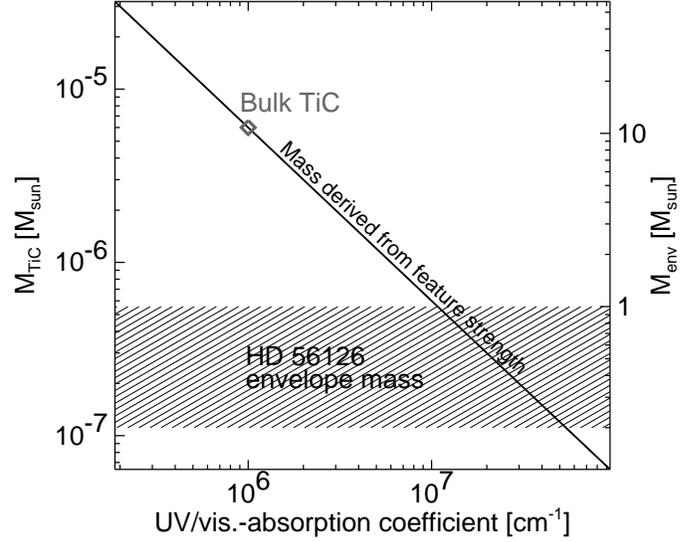}}
  \caption{The mass of the TiC component as a function of the
    UV/visible absorption coefficient ($\alpha$). We show the TiC mass
    derived from the strength of the ``21''~$\mu$m feature for
    different values of the UV/visible absorption coefficient (black
    line). A higher absorption coefficient requires less TiC. The grey
    diamond indicates the measured absorption coefficient of bulk TiC.
    The right abscissa shows the corresponding total envelope mass.
    The hatched area indicates reasonable values for the total
    envelope mass. The absorption coefficient of nano-TiC needs to be
    $\sim$20 times higher that bulk to be compatible with the observed
    strength of the ``21''~$\mu$m feature.}
  \label{hd56:fig:ticmass}
\end{figure}
We include MgS in the radiative transfer model using the optical
properties as described in Sect.~\ref{hd56:sec:mgs}. The only free
parameters are the MgS grain size and the relative mass fraction. We
find that a grain size of $a_\mathrm{MgS}\simeq$0.01$-$0.02~$\mu$m
fits best with the observed profile of the ``30''~$\mu$m feature.
Larger grains are too cold and cause the ``30''~$\mu$m feature to peak
at longer wavelength than observed. The mass of the MgS component is 4
per cent of the a-C, or 1.4\,10$^{-5}$\,M$_{\odot}$ at $d$=2.4 kpc;
this is in good agreement with the 9\,10$^{-6}$\,M$_{\odot}$ we derive
from the simple considerations above, assuming T$_\mathrm{MgS}$=150~K.

\subsection{Titanium carbide}
\label{hd56:sec:model:tic}
As discussed above (Sect.~\ref{hd56:sec:tic}) only TiC nano-crystals
are expected to show a strong ``21''~$\mu$m feature. Such
nano-crystals will undergo a temperature spike upon absorption of a
single UV/visible photon. Cooling then occurs through emission in the
``21''~$\mu$m feature. The mass of TiC is directly determined from the
energy balance: the IR flux is equal to the absorbed UV/visible flux,
which is given by our radiative transfer calculation. In
Fig.~\ref{hd56:fig:ticmass} we show the mass of TiC in the envelope
that we derive as a function of the UV/visible absorption coefficient
using the Ti/H abundance of 1.3\,10$^{-8}$
\citep{2000AnA...354..135V}. We also show the measured absorption
coefficient of bulk TiC. It is clear from Fig.~\ref{hd56:fig:ticmass}
that bulk TiC can not explain the strength of the feature even if all
the energy absorbed would be emitted solely in the ``21''~$\mu$m
resonance. In order to be compatible with the observed band strength
nano-TiC needs to absorb 20 times more efficiently per Ti-atom in the
0.2$-$0.7~$\mu$m wavelength range than bulk TiC.  This corresponds to
a visual absorption efficiency (Q$_\mathrm{vis}$) of $\sim$8 for a
10x10x10-atom nano-crystal. Such a high value for the visual
absorption coefficient is only possible if, due to the nanometer size
scale, a strong electronic resonance is present in nano-TiC. This
problem is already evident in the complete spectrum as presented in
Fig.~\ref{hd56:fig:sed}. Either the carrier of the ``21''~$\mu$m
feature is as abundant as MgS or it has a much higher UV/visible
absorptivity.
\begin{figure*}[!htp]
  \centerline{\includegraphics[width=18cm]{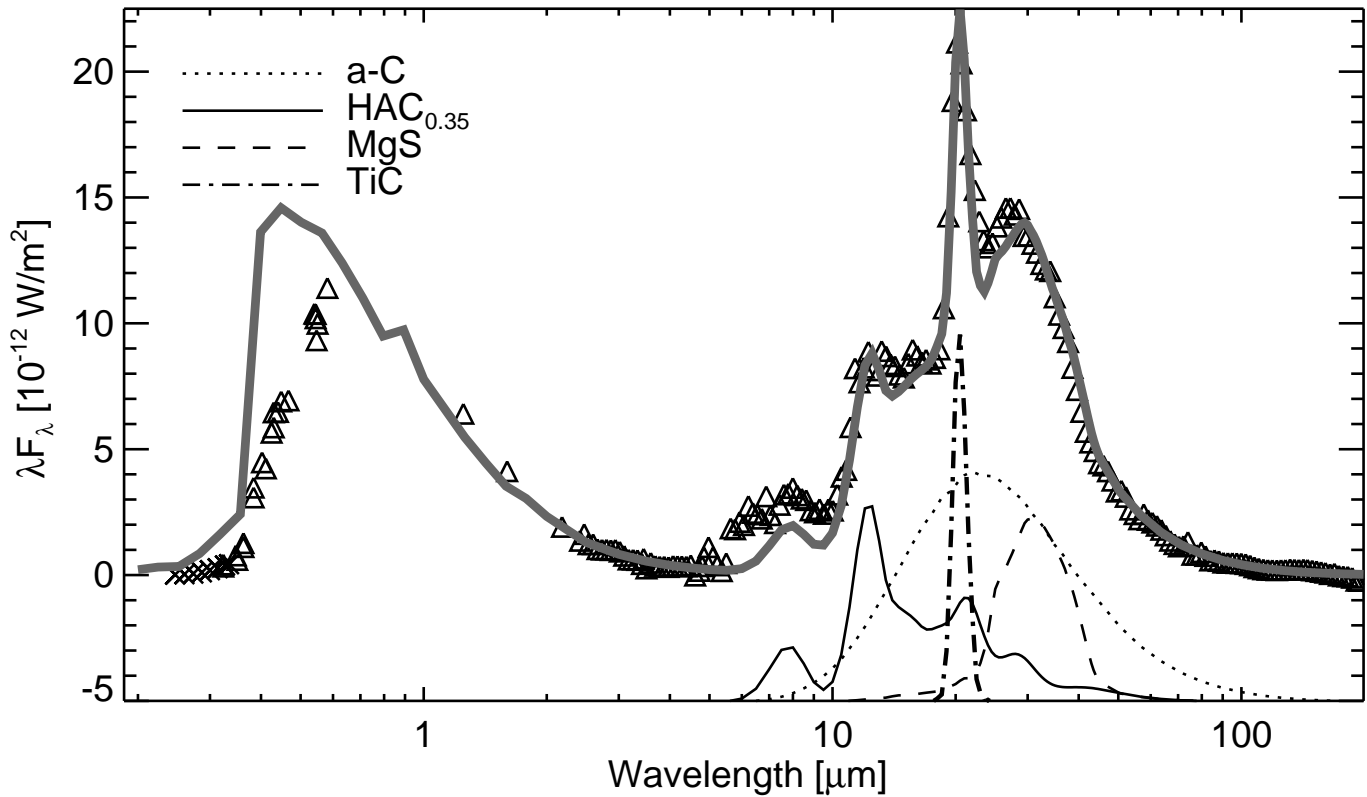}}
  \caption{A comparison between the observed SED (triangles) and our
    best fitting model (thick grey line). The model reproduces the
    dust continuum and the ``30''~$\mu$m feature very well. The model
    does not reproduce the strength of the 6$-$9~$\mu$m feature and
    the extinction in the UV. Underneath the IR emission spectrum we
    give an indication of the contributions of the various dust
    components offset by -5\,10$^{-12}$~W/m$^2$.}
  \label{hd56:fig:cbf}
\end{figure*}

\section{Results and discussion}
\label{hd56:sec:discussion}
\subsection{Best fit model}
\label{hd56:sec:best-fit-model}
\begin{figure}
  \centerline{\includegraphics[width=8.8cm]{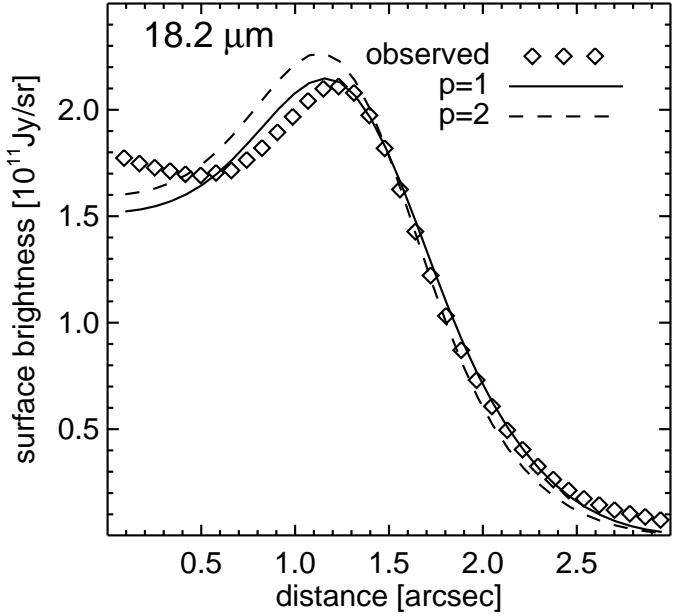}}
  \caption{ A comparison between the observed intensity profile
    (diamonds) as a function of distance from the centre of the
    nebula, the intensity profile from the best fitting radiative
    transfer model (line) and the profile predicted the $p$=2 model
    (dashed line). The observed profile is obtained by azimuthal
    averaging of the 18~$\mu$m image as published by
    \citet{2002ApJ...573..720K}. The model profiles are obtained by
    averaging the intensity profile of the model shown in
    Fig.~\ref{hd56:fig:cbf} and the $p$=2 model after convolution with
    a 2D-Gaussian PSF with the same full-width-at-half-maximum as the
    18~$\mu$m observation.}
  \label{hd56:fig:cbf_intensity}
\end{figure}
\begin{table}
  \caption{The best fitting model: parameters and derived quantities.} 
  \centerline{
    \begin{tabular}{l@{} l@{\ } l@{\ } r@{\ } c@{\ } l@{\ }}
      \hline
      \hline
      \multicolumn{5}{l}{$Central~star$}& $Comments$\\
      \hline
      $d$ &&[kpc] & 2.4 & &\\
      T$_\mathrm{eff}$ &&[K] & 7250 \\ 
      R$_{\star}$ & $\propto d^2$ & [R$_{\odot}$] & 49 \\ 
      L$_{\star}$ & $\propto d^2$ & [L$_{\odot}$] & 6000 \\ 
      \hline
      \multicolumn{6}{l}{$Dust~shell$}\\
      \hline
      R$_\mathrm{in}$  & $\propto d$ & [cm] & 4.5\,10$^{16}$ & & 1.2$^{\prime\prime}$ \\
      R$_\mathrm{out}$ & $\propto d$ & [cm] & 9.3\,10$^{16}$ & & 2.6$^{\prime\prime}$ \\
      v$_\mathrm{exp}$& & [km/s] & 10.7 &&\citet{1998ApJS..117..209K} \\
      $\Delta\mathrm{t}_\mathrm{in}$ & $\propto d$/v$_\mathrm{exp}$ & [yr] & 1300 \\
      $\Delta\mathrm{t}_\mathrm{out}$ & $\propto d$/v$_\mathrm{exp}$ & [yr] & 2800 \\
      $\rho_0$  && [g/cm$^3$] & 8\,10$^{-22}$ \\
      $p$&&&1&&$\rho$(r)=$\rho_0$\,$\left(\frac{\mathrm{r}}{\mathrm{R}_\mathrm{in}}\right)^{-p}$ \\
      \hline
      \multicolumn{3}{l}{$Dust~species$}&$Mass$&\%\\
      \hline
      dust         & $\propto d^2$ &  [M$_{\odot}$] & 7.4\,10$^{-4}$ & 100 \\
      a-C          & $\propto d^2$ &  [M$_{\odot}$] & 3.6\,10$^{-4}$ & 49 &
      $a$=0.01$-$1~$\mu$m\\
      HAC          & $\propto d^2$ &  [M$_{\odot}$] & 3.6\,10$^{-4}$ & 49 & H/(H+C)=0.35\\
      MgS          & $\propto d^2$ &  [M$_{\odot}$] & 1.4\,10$^{-5}$ & 2 & CDE\\
      TiC          & $\propto d^2$ &  [M$_{\odot}$] & 8\,10$^{-6}$    & $-$& Q$_\mathrm{vis}$=Q$_\mathrm{bulk}$\\
      TiC          & $\propto d^2$ &  [M$_{\odot}$] & 3\,10$^{-7}$    & $-$& Q$_\mathrm{vis}$=10\\
      \hline
      \multicolumn{6}{l}{$Envelope$}\\
      \hline
      M$_\mathrm{env}$ & $\propto d^2$ & [M$_{\odot}$] & 0.16& & M$_\mathrm{gas}$/M$_\mathrm{dust}$=220 \\
      $<$$\dot{M}$$>$& $\propto d$v$_\mathrm{exp}$ & [M$_{\odot}$/yr] & 1\,10$^{-4}$& & M$_\mathrm{gas}$/M$_\mathrm{dust}$=220 \\
      \hline
      \multicolumn{6}{c}{M$_\mathrm{gas}$/M$_\mathrm{dust}$=600, see Sect.~\ref{hd56:sec:disc:envel-mass}}\\
      \hline
      M$_\mathrm{env}$ & $\propto d^2$ & [M$_{\odot}$] & 0.44& & M$_\mathrm{gas}$/M$_\mathrm{dust}$=600 \\
      $<$$\dot{M}$$>$& $\propto d$v$_\mathrm{exp}$ & [M$_{\odot}$/yr] & 3\,10$^{-4}$& & M$_\mathrm{gas}$/M$_\mathrm{dust}$=600 \\
      \hline
      \hline
    \end{tabular}}
  \label{hd56:tab:model_params}
\end{table}
In Fig.~\ref{hd56:fig:cbf} we compare the predicted and observed SED.
Our best model fits the observations in many details. Note that the
ordinate scale is linear while previously published SED fits are
compared on a logarithmic basis. The peak of the SED, the shape and
strength of the ``30''~$\mu$m feature and the shape of the
``21''~$\mu$m feature are very well explained by the model. There are
two places where the model fails to reproduce the details of the
observations: {\it i}) The UV flux levels are over predicted.  It is
clear that an important UV absorption component is missing in our
model. Laboratory studies of the UV absorption properties of MgS and
TiC are therefore needed. {\it ii}) The strength of the 6$-$9~$\mu$m
band is too small in the model.  The discrepancy at 6$-$9~$\mu$m is
discussed in Sect.~\ref{hd56:sec:disc:T-hac}.

In Fig.~\ref{hd56:fig:cbf_intensity} we show the comparison between
the observed and predicted intensity profiles. The strength and
position of the intensity maximum and width of the nebula are
reproduced very satisfactorily by the $p$=1 model. In the outer parts
of the nebula this model under predicts the intensities slightly. This
may be due to deviations of the true PSF from the assumed Gaussian
PSF, some weak background in the observation or to weak nebular
emission that we do not include in the model. The best fitting $p$=2
model produces an intensity profile that is too much peaked towards
the inner edge of the dust shell.
  
Towards the centre of the circumstellar shell there is a rise in the
observed intensity, which is present in all available mid-IR images.
At 18.2 $\mu$m, this central region contains about 3 Jy which is
$\sim$100 times more than the photospheric contribution of the central
star at 18.2~$\mu$m.  Therefore, this rise must be due to the
departure from spherical symmetry of the nebula, causing more dust to
be present along the line of sight towards the central star or it is
due to an additional dust component which is not present in our model.
The nature and origin of such an additional dust component is at
present unknown.
  
We note that the emission from this region, which is possibly located
closer to the central star than the inner edge of the dust shell in
our model, cannot explain the plateau features and the high
temperatures they require (Sect.~\ref{hd56:sec:hac-temperature}),
because the flux contributions from this region are much too small and
the plateau features are emitted from a much more extended region,
e.g. Fig.~\ref{hd56:fig:timmi2} and the images presented in
\citet{1997ApJ...482..897M,1998ApJ...492..603D,2000ApJ...544L.141J,2002ApJ...573..720K}

We also note that our model fails to reproduce the observed intensity
profiles at shorter wavelengths (10.3, 11.7 and 12.5~$\mu$m). The
observed profiles are systematically broader than predicted. This is
an indirect but clear indication that the carriers of the plateau
emission features at those wavelengths are not in radiative
equilibrium, but are transiently heated to temperatures higher than
expected on the basis of our equilibrium model.

In Table~\ref{hd56:tab:model_params} we summarise the main parameters
of the best fit model. We find that the observations are best
explained with a compact dust shell that extents from
1.2$-$2.6$^{\prime\prime}$. The mass of the \emph{dust} contained in
the shell is $\sim$7\,10$^{-4}$~M$_{\odot}$. Assuming a constant
outflow velocity of 10.7~km/s \citep{1998ApJS..117..209K} we calculate
that the mass-loss commenced 2800~yr ago and halted 1300~yr ago. Using
a value of 220, as is often used for carbon-rich circumstellar matter,
for M$_\mathrm{gas}$/M$_\mathrm{dust}$ we find that the time averaged
mass-loss rate is $\sim$10$^{-4}$~M$_{\odot}$/yr.  As argued below, a
value of 600 for M$_\mathrm{gas}$/M$_\mathrm{dust}$ is more
appropriate for this low metallicity object, corresponding to a
mass-loss rate of $\sim$3\,10$^{-4}$~M$_{\odot}$. We find that 2 per
cent of the dust (by mass) is contained in MgS, consistent with the
observed photospheric abundance of Mg and S.

\subsection{Previous dust models}
\label{hd56:sec:disc:previouswork}
The circumstellar environment of HD~56126 has been studied before
using both mid-IR spectroscopic and imaging observations
\citep{1997ApJ...482..897M,1998ApJ...492..603D,2000ApJ...535..275H,2000ApJ...544L.141J}.
It is useful to compare the parameters we derive to those found
before.  One main difference between our model and previous work is
the fact that we treat the dust composition in considerable detail.
However, there are other significant differences as well.  We find
that the CSE of HD~56126 is well represented by a compact shell of
2.6$^{\prime\prime}$ radius. We have argued above that a larger outer
radius will cause an increase of the size of the nebula with
increasing wavelength, contrary to the observations.  Of course, a
preceding phase with a much lower mass-loss rate would go undetected.
The model of \citet{1995ApnSS.224..383S} contains a more extended
component. The absence of this more extended component in our model is
reflected in relatively low envelope mass of 0.16~M$_{\odot}$ that we
derive.  This is because cold material far away does not contribute to
the mid-IR fluxes but does contain mass.  This ``hidden'' component is
therefore important in determining the total envelope mass. Imaging at
sub-mm wavelengths is needed to better constrain the cold envelope
mass.

\citet{1995ApnSS.224..383S} find M$_\mathrm{env}$=0.54~M$_{\odot}$ at
a distance of 3~kpc, with two-thirds of the mass in the extended
component. The mass of the inner component that they derive
corresponds well to the mass of the a-C grains that we derive after
correcting for the distance.  \citet{1998ApJ...492..603D} derive an
envelope mass of 0.26~M$_{\odot}$ at 2.7~kpc, which corresponds to
0.2~M$_{\odot}$ at 2.4~kpc. These authors use a compact model with all
dust contained within 2.2$^{\prime\prime}$ but with two distinct fixed
temperatures.  In view of the difference in approach, the envelope
mass they find and our mass determination are compatible.
\citet{2000ApJ...535..275H} derive a mass-loss rate of
3.6\,10$^{-5}$~M$_{\odot}$/yr at the distance of 2~kpc, which
corresponds to 5\,10$^{-5}$~M$_{\odot}$/yr at 2.5~kpc, about half of
the value that we find.  \citet{2000ApJ...544L.141J} model the density
in the envelope as the result of a ``dying'' wind, i.e. a decreasing
mass-loss rate with time. This is qualitatively consistent with the
$\rho_0$\,(r/R$_\mathrm{in}$)$^{-1}$ density prescription that we use.
These authors find a maximum mass-loss rate of
$\sim$3\,10$^{-5}$~M$_{\odot}$/yr. However, they model the emission of
the dust with a simplified amorphous carbon opacity function. This
yields a mass estimate compatible with the warmest a-C grains
($a$=0.1~$\mu$m) in our model. Correcting for the mass of the other
grains, the different distance and the
M$_\mathrm{gas}$/M$_\mathrm{dust}$ ratio they use, they find
1.1\,10$^{-4}$~M$_{\odot}$/yr compatible with the mass-loss rate that
we derived.

\subsection{Envelope mass; M$_\mathrm{gas}$/M$_\mathrm{dust}$}
\label{hd56:sec:disc:envel-mass}
The total envelope mass that we and other authors derive scales with
the gas-to-dust mass ratio. For carbon-rich AGB and post-AGB stars
values between 200 and 250 are often used
\citep[e.g.][]{1986ApJ...303..327J,1997ApJ...482..897M} and in the
above derivations we have used 220. However this value is very
uncertain. Because the bulk of the dust is carbon-based one can expect
M$_\mathrm{gas}$/M$_\mathrm{dust}$ to strongly depend on the C/O
ratio. The photospheric abundances determination of
\citet{2000AnA...354..135V} yield [C/H]=8.53$-$8.77 and C/O=0.7$-$1.4.
Taking values in the high end of this range and assuming that all
carbon is locked in either CO (70 per cent) or dust (30 per cent) we
find M$_\mathrm{gas}$/M$_\mathrm{dust,C}\gtrsim$600, in which
M$_\mathrm{dust,C}$ is the mass of the carbon-based dust component.
This high value of M$_\mathrm{gas}$/M$_\mathrm{dust}$ follows directly
from the elemental abundance measurements. We find
M$_\mathrm{dust,C}$=7.4\,10$^{-4}$~M$_{\odot}$ (see
Table~\ref{hd56:tab:model_params}) which corresponds to
M$_\mathrm{gas}$=0.44~M$_{\odot}$. When we consider that the star had
a ZAMS-mass of $\sim$1.1~M$_{\odot}$ and the core-mass is roughly
0.6~M$_{\odot}$ we conclude that nearly the entire envelope was lost
during a high mass loss phase of $\sim$1500~yr. We also conclude that
only a small amount of the dust can be ``hidden'' in cold grains,
either far away from the star or in larger grains. This argues
strongly against the importance of grey extinction due to large
grains.  Using the envelope mass of 0.44~M$_{\odot}$ we find a high
time-averaged mass-loss rate of a $\sim$3\,10$^{-4}$~M$_{\odot}$. We
conclude that \emph{HD~56126 has experienced a short period
  (1500~year) of very strong mass loss (3\,10$^{-4}$~M$_{\odot}$/yr)
  during which the entire envelope was lost}. If we adopt lower C/H
and C/O values this conclusion would only be strengthened.

\subsection{The temperature of the HAC}
\label{hd56:sec:disc:T-hac}
One of the main discussion points concerning the relation between HAC
(or any other solid carrier) and the 3 to 17~$\mu$m features found in
C-rich environments is whether a solid particle \emph{in radiative
  equilibrium} is able to explain the observed spectra.  In particular
the large feature to continuum ratios found in the astronomical
spectra and the relatively short wavelengths (3~$\mu$m) at which
emission is found appear difficult to obtain in radiative equilibrium.
Indeed, one can immediately see from Fig.~\ref{hd56:fig:sed} precisely
that difficulty; the peak of the SED lies beyond 20~$\mu$m while
strong features on a weak continuum are present in the 5$-$17~$\mu$m
range.  The fundamental question is whether solid grains can reach the
temperature required to emit efficiently at these wavelengths.

We have tested this by building a detailed model of the optical
properties of HAC. We find that in radiative equilibrium HAC does not
reach a high enough temperature at the location of the dust. We have
tested this too for ``coal'' with the opacities as published by
\citet{1994AnA...285.1003G}, even though the main difference with HAC
is due to the interpolation these authors applied. We find that their
``coal'' reaches a maximum temperature of $\sim$350~K at the inner
edge of the dust shell. Which is not high enough to explain the
relative band-strength of the plateau features ($\sim$500$-$600~K).

This implies that the carriers that give rise to the plateau features
are not in radiative equilibrium with the stellar radiation field.
Most likely these features are due to very small dust grains or large
molecules that are transiently heated to high temperature. From the
PAH features that are present in the spectrum (especially at
3.3~$\mu$m) it is clear that a family of aromatic molecules is
present. The plateau features may well be carried by an ``extended PAH
family'' of aromatic molecules and clusters with attached aliphatic
sides groups.

\subsection{Magnesium sulfide}
\label{hd56:sec:disc:mgs}
We find that the amount of MgS required to explain the strength of the
``30''~$\mu$m feature is consistent with the measured abundances of Mg
and S. Taking into account the envelope mass that we derived above
(0.44~M$_{\odot}$ for M$_\mathrm{gas}$/M$_\mathrm{dust}$=600) we
conclude that only 25 per cent of the Mg- and S-atoms need to be
condensed into MgS to explain the strength of the feature. Because
HD~56126 is metal-poor this implies that in other sources that are
generally more metal-rich the strength of the ``30''~$\mu$m feature is
also consistent with the MgS identification.

\subsection{Titanium carbide}
\label{hd56:sec:disc:tic}
We have put TiC in our model. The optical properties of
nano-crystalline TiC are too poorly understood to accept or reject the
TiC identification for the ``21''~$\mu$m feature. We do note clearly
that the bulk optical properties of TiC are incompatible with the peak
position, feature-over-continuum ratio and strength of the
``21''~$\mu$m feature. The opacities of nano-TiC need to be $\sim$20
times higher in the visible than measured in the bulk in order to
explain the strength of the feature. It is interesting to note that a
very similar resonance has been found in other nano-crystalline metal
carbides, like VC, NbC \citep{vonHelden_VC,vanHeijnsbergen_NbC}.  Of
course, these species will not contribute much to the ``21''~$\mu$m
feature because of the much lower elemental abundance of Nb and V.
Better understanding the physics that gives rise to this resonance in
nano-carbides may give further insight into the identification of the
``21''~$\mu$m feature. We conclude that the question of the
identification of the ``21''~$\mu$m feature is still open and that
nano-crystalline metal carbides remain very interesting candidates.
Given the abundance of Fe and Ni, cubic iron-, nickel carbides
(Haxonite) may be good first candidates for future studies.

\section{Summary and conclusions}
\label{hd56:sec:summary-conclusions}
\begin{itemize}
\item We have presented a detailed study of the circumstellar envelope
  of the post-AGB star HD~56126. We present a large body of data from
  the literature, ISO spectroscopy and IR imaging. From a simple
  energy balance consideration we find a high effective temperature
  7250$\pm$250~K for the central star consistent with the temperature
  derived from detailed abundance analyses.
  
\item We have built a detailed dust radiative transfer model of the
  circumstellar envelope of HD~56126. To model the emission of the
  dust we use amorphous carbon (a-C), hydrogenated amorphous carbon
  (HAC), magnesium sulfide (MgS) and titanium carbide (TiC). We
  present a detailed parametrisation of the optical properties of HAC
  as a function of H/C content.
  
\item The radiative transfer model explains the ISO spectroscopy from
  2$-$200~$\mu$m in great detail.  The dust mass we derive is
  consistent with the dust mass as previously derived by other
  authors.  We find that the mid-IR imaging and spectroscopy is best
  reproduced by a single dust shell from 1.2 to 2.6$^{\prime\prime}$
  radius around the central star. This compact shell originates from a
  short period of very high mass-loss during which the mass-loss rate
  exceeded 10$^{-4}$~M$_{\odot}$/yr.
  
\item ``Classical'' HAC grains in radiative equilibrium with the local
  radiation field do not reach a high enough temperature to explain
  the strength of the 3.3$-$3.4 and 6$-$9~$\mu$m hydrocarbon features
  relative to the 11$-$17~$\mu$m hydrocarbon features. Hence we
  conclude that these features are carried by very small dust grains
  (or large molecules) which fluctuate in temperature upon absorption
  of a single UV/visible photon.
  
\item The strength of the ``30''~$\mu$m feature is compatible with its
  identification with MgS. This further strengthens the MgS
  identification of the ``30''~$\mu$m feature. The MgS temperature,
  derived from the observed profile allows an accurate determination
  of the MgS mass. We conclude that about 25 per cent of the Mg and S
  are locked up in MgS grains.
  
\item We find that the observed strength of the ``21''~$\mu$m feature
  poses a problem for the TiC identification. The low abundance of Ti
  requires very high absorption cross-sections in the UV and visible
  wavelength range to explain the strength of the feature. Laboratory
  studies of the UV absorption properties of TiC nano-crystals are
  therefore urgently needed. Other nano-crystalline metal carbides
  also exhibit similar resonances as found in nano-TiC and these metal
  carbides should also be investigated in the laboratory to clarify
  the nature of these resonances and the identification of the
  ``21''~$\mu$m feature.
\end{itemize}

\begin{acknowledgements}
  We thank Margaret Meixner and Mike Jura for the helpful discussions
  and kindly supplying their mid-IR images.  We thank Sun Kwok and
  Kevin Volk for kindly supplying the Gemini/OSCIR images we used in
  our analysis. We thank Hans van Winckel for providing his
  photometric data. We are very thankful to John Robertson and Uwe
  Kreibig for very useful directions, explanations and discussions on
  the properties of HAC and nano-metals, respectively. We thank
  Tsuneharu Koide for supplying us with the optical constants of TiC.
  SH and LBFMW acknowledge financial support from an NWO
  \emph{Pionier} grant (grant number 616-78-333).  This research has
  made use of the SIMBAD database, operated at CDS, Strasbourg,
  France.  This research has made use of NASA's Astrophysics Data
  System Bibliographic Services.  IA$^3$ is a joint development of the
  SWS consortium.  Contributing institutes are SRON, MPE, KUL and the
  ESA Astrophysics Division.
\end{acknowledgements}

\bibliographystyle{aa}
\bibliography{thesis}
\end{document}